\documentclass[aps,prl,reprint,superscriptaddress,amsmath,amssymb,nofootinbib,floatfix,longbibliography,raggedbottom]{revtex4-2}
\usepackage{amsmath,amssymb,bm,graphicx}
\usepackage[colorlinks=true,linkcolor=blue,citecolor=blue,urlcolor=blue]{hyperref}
\usepackage{orcidlink}

\hypersetup{pdfauthor={Pallab Basu, Vishnu Jejjala}}
\hypersetup{pdftitle={Bounds on entanglement fluctuations at holographic interfaces}}

\newcommand{\ceff}{c_\text{eff}}
\newcommand{\ccap}{c_\text{cap}}
\newcommand{\cLR}{c_{LR}}
\newcommand{\sst}{s_*}
\newcommand{\cR}{\mathcal R}
\newcommand{\cC}{\mathcal C}

\begin{document}

\title{Bounds on entanglement fluctuations at holographic interfaces}

\author{Pallab Basu\,\orcidlink{0000-0003-0006-7240}}
\email{pallab.basu@wits.ac.za}
\affiliation{Mandelstam Institute for Theoretical Physics, School of Physics, University of the Witwatersrand, Johannesburg, South Africa}
\affiliation{National Institute for Theoretical and Computational Sciences (NITheCS), Gauteng, South Africa}

\author{Vishnu Jejjala\,\orcidlink{0000-0003-2603-6717}}
\email{vishnu.jejjala@gmail.com}
\affiliation{Mandelstam Institute for Theoretical Physics, School of Physics, University of the Witwatersrand, Johannesburg, South Africa}
\affiliation{National Institute for Theoretical and Computational Sciences (NITheCS), Gauteng, South Africa}

\begin{abstract}
How small can entanglement fluctuations be at fixed entropy?
For a class of holographic conformal interfaces supported by a scalar with a potential, we prove $c_\text{eff}\leq c_\text{cap}\leq c_\text{eff}^2/c_{LR}$, relating entropy, capacity, and energy transmission.
The coupled gravitational response is a transport term minus a positive scalar contribution; the background Einstein equation bounds that subtraction.
A long Janus throat allows unbounded capacity at a finite entropy coefficient while the scalar response remains gapped.
\end{abstract}

\maketitle

\noindent\textit{Introduction.}\quad
Conformal interfaces, which join two two-dimensional conformal field theories (CFTs)
along a line while preserving conformal symmetry~\cite{Oshikawa:1996dj,Bachas:2001vj,Quella:2006de}, are characterized by universal data that include how much entanglement they allow across the interfaces and how much energy they transmit.
Let us call the two field theories CFT$_L$ and CFT$_R$.
The entanglement is measured by the vacuum entanglement entropy of a half line $A$
ending on the interface~\cite{Sakai:2008tt},
\begin{equation}
S_A=\frac{\ceff}{6}\,\ell+O(1)\,,\qquad \ell=\log\frac{L}{\epsilon}\,,
\label{eq:Seff}
\end{equation}
where $L$ and $\epsilon$ are infrared and ultraviolet cutoffs.
When $\mathrm{CFT}_L=\mathrm{CFT}_R$ and there is no interface, $\ceff=c$ and~\eqref{eq:Seff} reduces to the familiar half-line entropy of a homogeneous CFT.

Entropy alone does not fix the entanglement spectrum.
With $K_A=-\log\rho_A$ the modular Hamiltonian, $S_A=\langle K_A\rangle$ gives only the mean of the spectrum.
Its width is measured by the capacity of entanglement, $C_A=\langle K_A^2\rangle-\langle K_A\rangle^2$~\cite{Yao:2010woi,DeBoer:2018kvc}.
If the replica index is read as an inverse temperature, $C_A$ is the heat capacity of the spectrum~\cite{Hung:2011nu,Nakaguchi:2016zqi}; holographically, it is the rate at which a cosmic brane shrinks as its tension grows.
For the same half line,
\begin{equation}
C_A=\frac{\ccap}{6}\,\ell+O(1)\,.
\label{eq:Ccap}
\end{equation}
Without an interface, $\ccap=c$ as well, so the two coefficients agree.
Neither $\ceff$ nor $\ccap$ depends on the regularization scheme, so $\ccap$ is
a further piece of interface data, and an interface can distinguish it from $\ceff$.

It is natural to ask whether entropy and capacity obey any obvious general relation.
They do not: the density matrix $\rho_A=P/d$, with $P$ a rank-$d$ projector on the
Hilbert space of $A$, has $C_A=0$ while $S_A=\log d$ is arbitrary, and conversely
$C_A$ can be made arbitrarily large at fixed $S_A$.
A relation between $\ceff$ and $\ccap$ must therefore come from conformal dynamics.
Kusuki, Suzuyama, and Tang conjectured that every interface obeys $\ccap\geq\ceff$.
They checked the bound in free field theories, in holographic models where the
interface is a tension brane, and in defect perturbation theory. They also argued
that capacity carries information that neither entropy nor transmission
captures~\cite{Kusuki:2026arj}.

Kusuki, Suzuyama, and Tang also posed the question of extending the holographic bound to interfaces supported by bulk matter~\cite{Kusuki:2026arj}.
Varying the replica index then makes matter and geometry backreact on each other.
We compute this coupled response for a reflection symmetric interface supported
by a single monotonic canonical scalar with an arbitrary potential, subject to the
existence of the smooth wall backgrounds specified below.
At leading classical order we prove the conjectured lower bound
\begin{equation}
\boxed{\ceff\leq\ccap}\,.
\label{eq:bounds}
\end{equation}
For the same class of scalar walls, we obtain an upper bound controlled by energy transmission.
Denoting by $\cLR$ the energy transmission coefficient --- equal to the common central charge $c$ for a transparent interface~\cite{Quella:2006de,Meineri:2019ycm,Bachas:2022etu} --- we show
\begin{equation}
\ccap\leq\frac{\ceff^2}{\cLR}\,.
\label{eq:upper_bound}
\end{equation}
Within this class, the two bounds imply the sharpened transport constraint
\begin{equation}
\cLR\leq\frac{\ceff^2}{\ccap}\leq\ceff\,.
\label{eq:transport_bound}
\end{equation}
Equation~\eqref{eq:transport_bound} sharpens the holographic bound $\cLR\leq\ceff$ of~\cite{Karch:2024udk} (part of the proposed chain $\cLR\leq\ceff\leq\min(c_L,c_R)$, whose upper inequality was proved in~\cite{Karch:2023evr}).
Partially transmitting free fermion interfaces and weak interfaces generated solely by a marginal line deformation violate the capacity upper bound~\eqref{eq:upper_bound}, as shown in the Supplemental Material~\cite{Supplement}.
The value $\ceff^2/\cLR$ is realized, for example, in symmetric two-brane constructions~\cite{Kusuki:2026arj,Baig:2022cnb}; our analysis identifies the coupled scalar readjustment that lowers this ``transport value'' and the geometric budget that prevents the reduction from overshooting below $\ceff$.

\noindent\textit{The gravitational susceptibility.}\quad
Set the common asymptotic AdS radius to one and use
\begin{align}
I&=-\frac{1}{16\pi G}\int d^3x\sqrt g\,
 [R-(\partial\phi)^2-V(\phi)]+I_\text{boundary}\,,
\label{eq:action}\\
ds^2&=d\mu^2+f(\mu)(dv^2+\cosh^2v\,d\tau^2)\,,
\qquad f=e^{2A}\,.
\label{eq:geometry}
\end{align}
The coordinates are adapted to the entangling point.
On the boundary, $\tau=\log|x|$ is the dilatation about it, so its regulated range
is $\ell$; the limits $\mu\to+\infty$ and $\mu\to-\infty$ reach the half-planes of
$\mathrm{CFT}_R$ and $\mathrm{CFT}_L$, and $v\to\pm\infty$ reaches the interface.
Pure AdS is $f=\cosh^2\mu$.
The scalar $\phi_0(\mu)$ is odd and strictly increasing, and $f$ is even.
We assume that $f(0)$ is a global minimum, so the central geodesic computes the entropy of the chosen half line.
The boundary conditions fix the scalar sources and the physical metric on the quotient cylinder.
We allow normalizable tails in standard quantization with dimension $\Delta>1$, including marginal scalars, and the branch with vanishing logarithmic source at the Breitenlohner--Freedman bound $\Delta=1$.
We assume the existence and dominance of a saddle with replica and reflection symmetry that is differentiable at $n=1$.
The proof requires the Einstein--scalar dynamics and the stated boundary conditions in addition to the null energy condition.

The entropy is the length of the Ryu--Takayanagi curve: the central geodesic
$\mu=v=0$, which runs along $\tau$ from the entangling point.
Its length per unit $\tau$ is $\sqrt{f(0)}$, so $S_A=\tfrac{c}{6}\,\ell\,\sqrt{f(0)}$.
Together with the background Einstein equation and the transport
formula~\cite{Azeyanagi:2007qj,Gutperle:2015hcv,Bachas:2020yxv,Bachas:2022etu},
\begin{align}
\Phi\equiv\phi_0'^2&=\frac1f-A''>0\,,\qquad
\sst=\sqrt{f(0)}=\frac{\ceff}{c}\,,\nonumber\\
B\equiv\frac{c}{\cLR}&=\int_0^\infty\frac{d\mu}{f}
=1+\int_0^\infty\,d\mu\,\Phi\,,\qquad c=\frac{3}{2G}\,.
\label{eq:background}
\end{align}
Thus entropy measures the narrowest part of the geometry, whereas transmission depends on an integral across the wall.

The calculation of capacity requires the replica geometry. The replica manifold is the $n$-fold cover of the plane branched at the
entangling point; the conformal map $w=\log z$ turns it into a cylinder of
length $\ell$ with angular period $2\pi n$.
In the bulk, $\tau$ runs along the cylinder, while the replica angle winds around
the central geodesic $\mu=v=0$ in the $(\mu,v)$ plane.
The $\mathbb Z_n$ quotient therefore leaves a conical defect on that geodesic,
a cosmic brane of tension $(1-1/n)/(4G)$~\cite{Lewkowycz:2013nqa,Dong:2016fnf}.
Let $N_*(n)$ be its length per unit $\tau$.
By Dong's area formula, the brane's length is the entropy
$\widetilde S(n)=\tfrac{c}{6}\,\ell\,N_*(n)$ of
$\rho_A(n)=\rho_A^n/\operatorname{Tr}\rho_A^n$, the thermal state of the modular
Hamiltonian at inverse temperature $n$, which reduces to the Ryu--Takayanagi
length at $n=1$~\cite{Dong:2016fnf}.
For a thermal family, $\widetilde S'(n)=-C(n)/n$, where $C(n)$ is the variance of
the modular Hamiltonian of $\rho_A(n)$.
At $n=1$ this is the heat capacity of the introduction, $C_A=-\widetilde S'(1)$, and
therefore
\begin{equation}
\ceff=cN_*(1)\,,\qquad
\ccap=-c\,\partial_nN_*(n)\big|_{n=1}\,.
\label{eq:replica}
\end{equation}

Capacity thus asks: \emph{when the central geodesic acquires a small tension,
with the boundary data fixed, how much does its length change?}
The brane pulls on the geometry, opening a conical deficit around itself, and the
metric and the scalar wall rearrange in response.
Define the fractional lapse response by $g_{\tau\tau}=f\cosh^2v[1+2(n-1)h_\tau]+O((n-1)^2)$.
Then $N_*(n)=\sst\,[1+(n-1)h_\tau(0,0)]+O((n-1)^2)$, giving
\begin{equation}
\frac{\ccap}{\ceff}=-h_\tau(0,0)\,.
\label{eq:question}
\end{equation}
In pure AdS, $N_*=1/n$ and $\ccap=\ceff=c$.
Capacity is in this sense a gravitational susceptibility: how much the brane
shrinks per unit tension.
At leading semiclassical order it is also the variance of the area in units
of $4G$~\cite{DeBoer:2018kvc,Dong:2018seb}, so~\eqref{eq:bounds} places a
lower bound on the universal logarithmic coefficient of the area variance with
the coefficient of the mean area held fixed.

To compute the response we work in uniform scalar gauge: the monotonic scalar
is the radial coordinate, $\phi=\phi_0(\mu)$ is unperturbed, and the scalar
response is carried by the metric through $g_{\mu\mu}=1+2(n-1)\eta$.
The linearized Einstein--scalar constraints then reduce to a single equation for
$\zeta=f\Phi\eta$, in which the conical defect appears as a point source at the
tip, $\eta\simeq-\tfrac12\log(\mu^2+\sst^2v^2)$.
The background equations absorb the derivatives of the potential into $f$ and
$\Phi$, so no independent $V''$ term appears and the operator is manifestly
positive.
Expanding in even eigenfunctions of $\Delta_2=\partial_v^2+\tanh v\,\partial_v$,
the radial Green function obeys
\begin{equation}
\left[-\partial_\mu\frac1\Phi\partial_\mu+
\frac{k_p^2}{f\Phi}\right]G_p(\mu,\mu')=\delta(\mu-\mu')\,,
\quad k_p^2=p^2+\frac94\,,
\label{eq:green}
\end{equation}
with the decaying solution selected by the fixed scalar sources.

Solving for $h_\tau(0,0)$ takes two steps.
Integrating the constraints along the axis $v=0$ from the boundary, where the physical boundary conditions determine $h_\tau$, to the tip gives
\begin{equation}
\begin{gathered}
\frac{\ccap}{\ceff}=\sst+\int_0^\infty d\mu\,\Xi_\Phi(\mu)\,\eta(\mu,0)\,,\\
\Xi_\Phi(\mu)=\int_\mu^\infty\Phi(\mu')\,d\mu'\,,
\end{gathered}
\label{eq:axis}
\end{equation}
so the brane length responds to the radial stretching along the axis, weighted
by the scalar gradient energy outside each point.
The stretching $\eta$ is the response to the tip source.
The radial operator in~\eqref{eq:green} maps $\Xi_\Phi(|\mu|)$ to the weighted source $k_p^2\Xi_\Phi(|\mu|)/(f\Phi)$ plus $2\delta(\mu)$.
Each mode therefore splits into a part fixed by $\Xi_\Phi(0)=B-1$ and a Green function remainder, giving
\begin{equation}
\frac{\ccap}{c}=\sst^2(B-\mathcal J)\,,\qquad
\mathcal J=2\pi\int_0^\infty dp\,
\frac{\rho(p)}{k_p^2}G_p(0,0)\geq0\,,
\label{eq:response}
\end{equation}
where $\rho(p)$ is the positive AdS$_2$ spectral measure specified in the Supplemental Material~\cite{Supplement}.
The two terms have distinct origins.
Holding the boundary arc fixed, the defect changes the conformal length of every
constant scalar slice by $\pi\sst\int_\mu^\infty d\mu'/f$; at the tip this gives
the transport term $c\sst^2B=\ceff^2/\cLR$.
The second term is the reduction due to the scalar wall's relaxation.
Since $G_p(0,0)>0$, $\mathcal J\geq0$ and the upper bound follows at once.
The lower bound is the real question: how large can the reduction be?

\noindent\textit{Why the scalar response is bounded.}\quad
By~\eqref{eq:response}, $\ccap\geq\ceff$ is equivalent to $\mathcal J\leq B-1/\sst$,
and the right-hand side is fixed by the background.
With $\cC=A'^2+1/f$ and $T=A'/\sqrt{\cC}$, the Einstein equation gives
$\cC'=-2A'\Phi$; integrating from the tip ($\sqrt{\cC}=1/\sst$) to the
boundary ($\sqrt{\cC}=1$),
\begin{equation}
B-\frac1{\sst}=\int_0^\infty d\mu\,\Phi(1-T)\,.
\label{eq:budget}
\end{equation}
Since $|T|<1$, this alone recovers $\cLR\leq\ceff$~\cite{Karch:2024udk}.
The scalar gradient energy that supports the wall sets a budget, and the
lower bound says the replica response stays within it.

To compare the two, we bound the Green function from above.
For any locally absolutely continuous trial with $\alpha_p(0)=1$, finite weighted energy, and $\alpha_p(\mu)G_p(\mu,0)\to0$ at infinity, integration by parts and Cauchy--Schwarz give
\begin{equation}
G_p(0,0)\leq\frac12\int_0^\infty d\mu\,
\Phi\left(\alpha_p^2+\frac{f}{k_p^2}\alpha_p'^2\right)\,,
\label{eq:variational}
\end{equation}
with equality for $\alpha_p=-2\Phi^{-1}\partial_\mu G_p(\mu,0)$.
It therefore suffices to exhibit one trial whose energy density, summed over $p$
with the weight in~\eqref{eq:response}, is at most $1-T$ pointwise.

Such trials come from an exactly solvable kernel,
\begin{equation}
w_z(v)=\int_z^1\frac{du\,u^2}{\sinh^2v+u^2}\,,
\qquad 0\leq z\leq1\,,
\label{eq:kernel}
\end{equation}
with $(2-\Delta_2)w_0=\pi\delta(v)$.
In pure AdS the profiles $\widehat\alpha_p(m)=k_p^2(w_{\tanh m})_p/\pi$ have summed energy $Z(z)+H(z)=1-z$, with $z=\tanh m$, $Z(z)=1-3z/2+z^3/2$, and $H(z)=z(1-z^2)/2$.
On a general wall, define the running maximum
\begin{equation}
\begin{gathered}
m(\mu)=\max_{0\leq\nu\leq\mu}\operatorname{artanh}T(\nu)\,,
\\
U(\mu)=\tanh m(\mu)\geq T(\mu)\,,
\end{gathered}
\label{eq:running_maximum}
\end{equation}
and set $\alpha_p(\mu)=\widehat\alpha_p(m(\mu))$.
Where the running maximum increases, the Einstein equation gives $\upsilon\equiv m'/\sqrt{\cC}=1-\Phi/\cC\in[0,1]$.
On its plateaus $m'=0$ and $\upsilon=0$.
The summed trial energy density is $\mathcal K_\alpha=Z(U)+\upsilon^2H(U)$, so
\begin{equation}
(1-T)-\mathcal K_\alpha
=(U-T)+(1-\upsilon^2)H(U)\geq0\,.
\label{eq:comparison}
\end{equation}
Integrating against $\Phi$ and using~\eqref{eq:budget} and~\eqref{eq:variational} proves~\eqref{eq:bounds}.
The gap $\ccap-\ceff$ is a sum of nonnegative terms and is strictly positive for
nontrivial walls; pure AdS ($\Phi=0$) saturates both bounds~\cite{Supplement}.

\begin{figure*}[t]
\includegraphics[width=\textwidth]{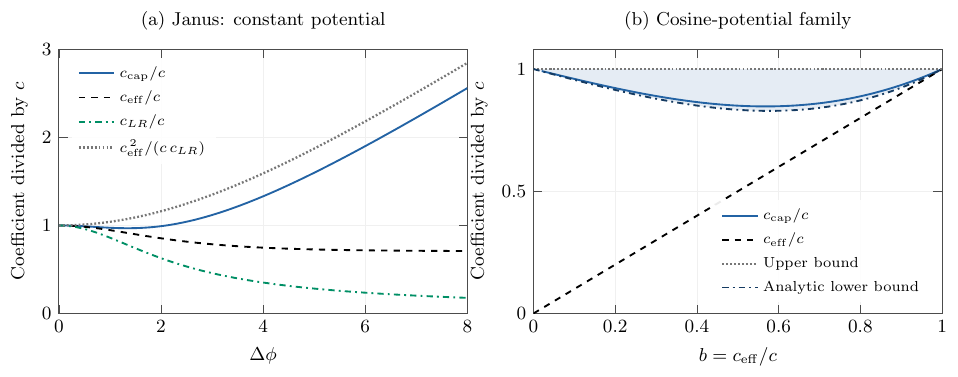}
\caption{\textsf{
Capacity probes the response of an interface to a conical defect.
(a) In Janus, capacity (solid blue) eventually grows without bound, while entropy (dashed black) approaches $c/\sqrt2$ and transmission (dash-dotted green) vanishes.
The dotted gray curve is the transport upper bound.
(b) In the family with a cosine potential, $\ceff/c=b$ and $\cLR/c=b^2$, but $\ccap/c\to1$ as $b\to0$.
The shaded region is the analytic interval in~\eqref{eq:cosine_bound}.
Panel (b) traces a family of bulk potentials, with $b\to0$ taken as a classical limit.
The curves evaluate the response formula~\eqref{eq:response}.}}
\label{fig:profiles}
\end{figure*}

\noindent\textit{What the response reveals.}\quad
Two examples illustrate how the scalar response separates capacity from entropy.
Their asymptotic limits concern the classical logarithmic coefficients, with the replica derivative taken first at fixed model parameters.
For the Janus interface~\cite{Bak:2003jk,Bak:2007jm}, $V=-2$ and $f=(1+\kappa\cosh2\mu)/2$, where $\kappa=\sqrt{1-2\gamma^2}$ and $\phi_0'=\gamma/f$.
The coupling jump is $\Delta\phi=\sqrt2\operatorname{artanh}(\sqrt2\gamma)$.
Capacity first decreases as $\ccap/c=1-\gamma^2/6+O(\gamma^4)$, then increases and exceeds $c$ near $\Delta\phi=2.10$ [Figure~\ref{fig:profiles}(a)].
The supplement uses polar coordinates to check the next perturbative coefficient independently.

The behavior for a large coupling jump has a simple geometric origin.
As $\kappa\to0$, an increasingly long throat develops with $f\simeq1/2$ and $\Phi\simeq2$.
It makes $B=\int_0^\infty d\mu/f$ large while leaving the central geodesic length per unit logarithmic coordinate finite.
The scalar operator tends to $-\tfrac12\partial_\mu^2+k_p^2$, which remains gapped, and its subtraction in~\eqref{eq:response} approaches a finite constant.
Consequently,
\begin{align}
\frac{\ceff}{c}&\longrightarrow\frac1{\sqrt2}\,,\qquad
\frac{\cLR}{c}\sim\frac{\sqrt2}{\Delta\phi}\,,\nonumber\\
\frac{\ccap}{c}&=\frac{\Delta\phi}{2\sqrt2}-0.287536\ldots+o(1)\,.
\label{eq:janus_limit}
\end{align}
The growing susceptibility comes from the increasingly long throat: boundary normalization amplifies the geometric term while the scalar contribution stays finite.

A second example tests the role of a nonconstant potential.
For $V=-2-\alpha_b^2\cos^2(\phi/\alpha_b)$, with $\alpha_b=\sqrt{1-b^2}/b$ and $0<b<1$, an exact wall is $f=b^2\cosh^2\mu$, $\phi_0=\alpha_b\arctan(\sinh\mu)$.
It has $\ceff=bc$, $\cLR=b^2c$, and an analytic radial Green function.
A direct estimate gives
\begin{equation}
1-\frac49b(1-b^2)<\frac{\ccap}{c}<1\,.
\label{eq:cosine_bound}
\end{equation}
Thus the normalized entropy coefficient tends to zero while the capacity coefficient tends to one [Figure~\ref{fig:profiles}(b)].
Varying $b$ traces a family of bulk potentials, each with a smooth wall for $b>0$.
The endpoint is taken after the classical large-$c$ limit; the Supplemental Material~\cite{Supplement} records the semiclassical conditions and limiting procedures.

\noindent\textit{An independent CFT check.}\quad
Near the transparent interface, conformal perturbation theory distinguishes a jump in a bulk coupling from an interaction localized on the interface.
Consider an analytic centered interface branch whose first-order deformation is a pure jump of an exactly marginal bulk coupling, with endpoint interchange acting by the adjoint~\cite{Komatsu:2025cai}.
Using the replica continuation specified in the supplement, we find
\begin{equation}
\frac{\ceff}{c}=1-\frac38\cR+O(\cR^2)\,,\qquad
\frac{\ccap}{c}=1-\frac14\cR+O(\cR^2)\,,
\label{eq:cft}
\end{equation}
where $\cR=1-\cLR/c$.
These slopes occur for the free boson in~\cite{Kusuki:2026arj}; the calculation extends them to small bulk coupling jumps along the stated interface branches.
For Janus, $\cR=2\gamma^2/3+O(\gamma^4)$, so~\eqref{eq:cft} reproduces both gravitational coefficients.
An independent marginal line perturbation instead gives $\ccap/c=1+\cR/2+\cdots$~\cite{Kusuki:2026arj}.
The location of the interaction therefore matters even at weak reflection.

\noindent\textit{Outlook.}\quad
The result identifies a bulk mechanism for the conjectured lower bound on entanglement fluctuations.
Boundary normalization fixes a transport contribution to the gravitational susceptibility, scalar readjustment reduces it, and the background Einstein equation limits that reduction.
This proves the bound for arbitrary potentials admitting the specified class of walls and replica branches.
The response formula and geometric comparison provide a starting point for extensions to multiple scalars, asymmetric interfaces, and quantum corrections.
Reflection symmetry fixes equal tip jumps on the two sides and supplies a common normalization for the geometric comparison.
An asymmetric extension requires matching two independent boundary responses and constructing a corresponding two-sided estimate for the Green function.
The Janus throat demonstrates the physical distinction between entropy and capacity: a finite entropy coefficient can coexist with arbitrarily large fluctuations of modular energy.

\noindent\textit{Acknowledgments.}\quad
We thank Joydeep Naskar for conversations.
VJ is supported by the South African Research Chairs Initiative of the Department of Science, Technology, and Innovation and the National Research Foundation (grant 78554).
Claude (Anthropic) and ChatGPT in Codex (OpenAI), used in November 2026, assisted with derivations, numerical checks, and manuscript preparation.
The Supplemental Material and reproduction archive document the separate analytical and numerical checks.

\noindent\textit{Data availability.}\quad
The \href{https://github.com/pallabbasu/capacity}{reproduction archive} contains the numerical scripts, reference figure data, and spectral and polar outputs, including all seven cases of the quartic convergence study.

\nocite{Bachas:2020yxv,Bachas:2022etu,Karch:2023evr,Hung:2011nu,Dong:2016fnf,Nakaguchi:2016zqi,Skenderis:2002wp,Baig:2022cnb,Kusuki:2026arj,Karch:2024udk,Balthazar:2022hzb,Mollabashi:2021xsd,Witten:2018xfj,Sakai:2008tt,Eisler:2012xry,Meineri:2019ycm,Komatsu:2025cai,Karch:2018uft}

\bibliography{refs}

\end{document}


\title{Supplemental Material for\texorpdfstring{\\}{: } Bounds on entanglement fluctuations at holographic interfaces}

\author{Pallab Basu}
\email{pallab.basu@wits.ac.za}
\affiliation{Mandelstam Institute for Theoretical Physics, School of Physics, University of the Witwatersrand, Johannesburg, South Africa}
\affiliation{National Institute for Theoretical and Computational Sciences (NITheCS), Gauteng, South Africa}

\author{Vishnu Jejjala}
\email{vishnu.jejjala@gmail.com}
\affiliation{Mandelstam Institute for Theoretical Physics, School of Physics, University of the Witwatersrand, Johannesburg, South Africa}
\affiliation{National Institute for Theoretical and Computational Sciences (NITheCS), Gauteng, South Africa}

\maketitle

The code we used in this work is available on \href{INSERT URL}{github}.

\section{Replica conventions and background equations}\label{sec:one}
The calculation separates into three steps.
First we solve the linear response of the scalar wall to the replica defect, keeping the physical boundary geometry and scalar sources fixed.
Second we identify the resulting change in the defect length with the capacity.
Finally we bound the scalar response by comparing a variational trial with the background Einstein equation.
Sections~\ref{sec:six} and~\ref{sec:seven} evaluate the response in two families, and Section~\ref{sec:eight} provides a CFT check near the transparent interface.

Our Euclidean action and background conventions are
\begin{align}
I&=-\frac{1}{16\pi G}\int d^3x\sqrt g\,
       [R-(\partial\phi)^2-V(\phi)]+I_\text{boundary}\,,
\label{sm:action}\\
ds^2&=d\mu^2+f(\mu)[dv^2+\cosh^2v\,d\tau^2]\,,
\qquad f=e^{2A}\,,\qquad c=\frac{3}{2G}\,,
\label{sm:background}\\
\Phi&\equiv\phi_0'^2=\frac1f-A''\,,\qquad
V(\phi_0)=-\frac1f-A''-2A'^2\,.
\label{sm:background_equations}
\end{align}
The common asymptotic AdS radius is one.
The coordinate $\tau$ is the translation direction whose regulated length gives the logarithm $\ell=\log(L/\epsilon_\text{UV})$.
It is not the Euclidean time of the original plane.
We assume that $f$ is even, $\phi_0$ is odd, and $\phi_0'>0$ at every finite $\mu$.
We also assume that $f(0)$ is a global minimum, so the central extremal line computes the entropy of the half line.
The scalar boundary conditions are specified in Section~\ref{sec:three}.
We use the differentiable replica branch specified in Section~\ref{sec:nine}, which collects the scope and physical limitations of the results.

Define
\begin{equation}
\sst=\sqrt{f(0)}\,,\qquad
\ceff=c\sst\,,\qquad
B\equiv\frac{c}{\cLR}=\int_0^\infty\frac{d\mu}{f(\mu)}
=1+\int_0^\infty\Phi\,d\mu\,.
\label{sm:observables}
\end{equation}
The last equality follows by integrating~\eqref{sm:background_equations}, using $A'(0)=0$ and $A'(\infty)=1$.
The transport identification follows from~\cite{Bachas:2020yxv,Bachas:2022etu}.
For an interface joining CFTs of central charges $c_L$ and $c_R$, the general entropy bound is $\ceff\le\min(c_L,c_R)$~\cite{Karch:2023evr}; the transmission bound $\cLR\le\ceff$ is established holographically~\cite{Karch:2024udk}.
We will prove, at leading classical order,
\begin{equation}
\ceff\le\ccap\le\frac{\ceff^2}{\cLR}\,.
\label{sm:bound}
\end{equation}
Both inequalities are strict for a nontrivial smooth wall in the stated class.
The transparent interface saturates both and is treated directly as pure AdS, since formulas containing $\Phi^{-1}$ are singular at $\Phi=0$.

Let $F(n)=\ln\operatorname{Tr}\rho_A^n=\ell[e(n)-ne(1)]+O(1)$, where $-e(n)$ is the ground state energy on the replica circle of circumference $2\pi n$.
With $K=-\log\rho_A$, the entropy and capacity are $S_A=-F'(1)$ and $C_A=F''(1)=\langle K^2\rangle-\langle K\rangle^2$.
The normalization is $S_A=(\ceff/6)\ell+O(1)$ and $C_A=(\ccap/6)\ell+O(1)$.
The cosmic brane has tension $(1-1/n)/(4G)=(n-1)/(4G)+O((n-1)^2)$ and length density $N_*(n)$, so
\begin{equation}
\frac{\ceff}{6}=e(1)-e'(1)\,,\qquad
\frac{\ccap}{6}=e''(1)\,,\qquad
N_*(n)=\frac6c[e(n)-ne'(n)]\,.
\label{sm:g:replica}
\end{equation}
These expressions use the same physical regulator and fixed scalar sources for every replica index.
They combine the thermodynamic and cosmic brane descriptions of holographic R\'enyi entropy~\cite{Hung:2011nu,Dong:2016fnf,Nakaguchi:2016zqi}.

The brane length has a direct entropic meaning.
The normalized state $\rho_A(n)=\rho_A^n/\operatorname{Tr}\rho_A^n$ is thermal for the modular Hamiltonian at inverse temperature $n$, and its entropy is
\begin{equation}
\widetilde S(n)=F-nF'=\ell\,[e(n)-ne'(n)]=\frac c6\,\ell\,N_*(n)\,,
\label{sm:g:escort_area}
\end{equation}
the area of the cosmic brane in units of $4G$~\cite{Dong:2016fnf}.
Since $\widetilde S'(n)=-nF''(n)=-C(n)/n$, with $C(n)$ the variance of the modular Hamiltonian of $\rho_A(n)$ (Section~\ref{sec:eight}), the rate at which the brane shrinks at $n=1$ is the heat capacity of the entanglement spectrum, $C_A=-\widetilde S'(1)$.
For a flat spectrum $\rho_A(n)=\rho_A$, the brane length is independent of $n$ and $C_A=0$.
For the renormalized energy variable $M(n)=12e(n)/(cn)$ and $y=n^{-2}$, direct differentiation gives
\begin{equation}
\frac{dM}{dy}=nN_*(n)\,,\qquad
\ccap-\ceff=2c\,\frac{d^2M}{dy^2}\bigg|_{y=1}\,.
\label{sm:g:energy_convexity}
\end{equation}
The capacity bound is therefore local convexity of this energy variable at $n=1$.
It is not implied by positivity of the capacity alone.

For $ds^2=N^2d\tau^2+\gamma_{ab}dx^adx^b$, the Einstein and scalar equations are
\begin{align}
2K&=V+|D\phi|^2\,,\qquad \Delta_\gamma N=-VN\,,\nonumber\\
D_aD_bN&=N\left[\tfrac12(|D\phi|^2-V)\gamma_{ab}-\partial_a\phi\partial_b\phi\right]\,,\qquad
D_a(ND^a\phi)=\tfrac12NV'\,,
\label{sm:g:static}
\end{align}
where $K$ is the Gaussian curvature of $\gamma$.
The geometric mass $\mathcal M=N^2-|DN|^2$ obeys the following identity on the reflection axis $v=0$, directed from the replica tip toward the boundary:
\begin{equation}
d\mathcal M=\tfrac12(|D\phi|^2+V+2)\,d(N^2)\,.
\label{sm:mass_identity}
\end{equation}
For normalizable scalar falloff faster than $e^{-\mu}$, its boundary limit equals $M(n)$ and
\begin{equation}
N_*^2=M(n)-\mathcal I_n\,,\qquad
\mathcal I_n=\frac12\int_{v=0}(|D\phi|^2+V+2)\,d(N^2)\,.
\label{sm:g:identity}
\end{equation}
Slower falloff can contribute finite scalar terms to the renormalized boundary energy.
We do not identify the renormalized boundary energy with the bare mass in that case: the physical boundary derivation in Section~\ref{sec:three} establishes the response formula directly.

\section{The coupled scalar and metric response}\label{sec:two}
The uniform scalar perturbation is
\begin{equation}
ds^2=(1+2\delta\eta)d\mu^2+f(1+2\delta h_v)dv^2
+f\cosh^2v(1+2\delta h_\tau)d\tau^2\,,\qquad \delta=n-1\,,\qquad\phi=\phi_0(\mu)\,.
\label{sm:g:pert}
\end{equation}
For the metric \eqref{sm:g:pert} with $\phi=\phi_0(\mu)$, $q\equiv f'/f$, $t_v\equiv\tanh v$, and $\Delta_2=\partial_v^2+t_v\partial_v$, the first-order part of $E_{\mu\nu}\equiv R_{\mu\nu}-\partial_\mu\phi\,\partial_\nu\phi-Vg_{\mu\nu}$ is
\begin{align}
E_{\mu\mu}&=-(h_\tau+h_v)''-q(h_\tau+h_v)'+q\,\eta'-f^{-1}\Dt\eta+(f''+2)\,\eta/f\,,\\
E_{\mu v}&=-\partial_\mu\partial_v h_\tau-t_v\,\partial_\mu(h_\tau-h_v)+\tfrac12q\,\partial_v\eta\,,\\
\frac{E_{\tau\tau}}{f\cosh^2v}-\frac{E_{vv}}{f}&=-(h_\tau-h_v)''-q(h_\tau-h_v)'+f^{-1}(\partial_v^2-t_v\partial_v)\eta\,,\\
f\Big[E_{\mu\mu}-\frac{E_{vv}}{f}-\frac{E_{\tau\tau}}{f\cosh^2v}\Big]&=2\big(\partial_v^2h_\tau+2t_v\partial_v h_\tau-t_v\partial_vh_v-2h_v\big)+f'(h_\tau+h_v)'-(f''-2)\,\eta\,,
\label{sm:linearized_components}
\end{align}
where primes are $\mu$ derivatives and the background relations $\Phi=1/f-A''$, $V=-(1/f+A''+2A'^2)$ were used.
They follow by expanding the full Ricci tensor and substituting the background equations.
The gauge leaves the freedom $v\to v+\delta\,\xi(v)$, under which $h_v\to h_v+\partial_v\xi$, $h_\tau\to h_\tau+t_v\xi$ and $\eta$ is invariant; $h_\tau$ on the axis $v=0$ is invariant.

Write $h_v=\psi+\partial_v^2E$ and $h_\tau=\psi+t_v\partial_vE$.
With $\chi=2\psi+\Delta_2E$, the four combinations above become
\begin{align}
E_{\mu\mu}&=-\chi''-q\chi'+q\eta'-f^{-1}\Delta_2\eta+(f''+2)\eta/f\,,
\nonumber\\
E_{\mu v}&=-\partial_v[\psi'+E'-\tfrac12q\eta]\,,
\nonumber\\
\frac{E_{\tau\tau}}{f\cosh^2v}-\frac{E_{vv}}f
&=(\partial_v^2-t_v\partial_v)[E''+qE'+\eta/f]\,,
\nonumber\\
f\left[E_{\mu\mu}-\frac{E_{vv}}f-\frac{E_{\tau\tau}}{f\cosh^2v}\right]
&=2(\Delta_2-2)\psi+f'\chi'-(f''-2)\eta\,.
\label{sm:constraint_combinations}
\end{align}
The equations of motion are therefore
\begin{equation}
(fE')'=-\eta\,,\qquad \psi'=\tfrac12q\,\eta-E'\,,\qquad (\Dt-2)\,\Psi=-2f\Phi\,\eta\,,\qquad \Psi\equiv2\psi+f'E'\,,
\label{sm:linearized_constraints}
\end{equation}
where $f''-f'^2/f-2=-2f\Phi$ was used, and they imply $\Psi'=-2f\Phi E'$.
Eliminating $E$ gives $(\Psi'/\Phi)'+(\Dt-2)\Psi/(f\Phi)=0$, and applying $\Dt-2$ gives the same equation for $\zz=f\Phi\eta$, which is~\eqref{sm:master} away from the source.
For $V=-2$, where $\Phi=\gamma^2/f^2$, it reduces to $[\partial_\mu^2+f^{-1}\Dt-4]\eta=0$.
Adding $E_{\mu\mu}$ to the third combination gives
\begin{equation}
(fh_\tau')'=\tfrac12(f'\eta)'+\eta-t_v\partial_v\eta\,,
\label{sm:g:Daxis}
\end{equation}
which on the axis integrates to $fh_\tau'=\tfrac12f'\eta+\int_0^\mu\eta$.

\subsection{Source and boundary conditions}
On the smooth $n$-fold cover $\phi-\phi(p)\propto\text{Re}\,w^n$ near the fixed point, so in the quotient the metric is $|\tilde w|^{2(1/n-1)}|d\tilde w|^2$ with $\tilde w\propto\mu+i\sst v$.
Hence $\eta\simeq-\tfrac12\ln(\mu^2+\sst^2v^2)$ and $\zz$ has the source $2\pi\sst\,\delta(\mu)\delta(v)$ in~\eqref{sm:master}, while $h_\tau$ is regular.
The complete equation is therefore
\begin{equation}
\left[-\partial_\mu\frac1\Phi\partial_\mu
 +\frac{2-\Dt}{f\Phi}\right]\zz
=2\pi\sst\,\delta(\mu)\delta(v)\,,
\qquad \zz=f\Phi\eta\,,\qquad
\Dt=\partial_v^2+\tanh v\,\partial_v\,.
\label{sm:master}
\end{equation}
At the conformal boundary the fixed scalar source selects $\zz\to0$ and $\eta=O(e^{-2\mu})$, as shown below.
Toward the interface endpoints $|v|\to\infty$, $\eta\to0$ and the physical lapse perturbation remains bounded.
The auxiliary fields $\psi$ and $E$ need not separately obey this last condition.
Uniform scalar gauge uses the monotonic background scalar as a radial coordinate.
The derivatives of the scalar potential are consequently absorbed into $f$ and $\Phi$ through the background equations, rather than appearing as an independent $V''$ term in~\eqref{sm:master}.

\section{Boundary conditions and the exact capacity formula}\label{sec:three}

\subsection{Scalar sources and endpoint regularity}
We work on the branch with reflection symmetry and an odd, monotone scalar profile and hold its two asymptotic sources fixed.
On the right asymptotic end, write
\begin{equation}
f\sim a_\infty^2e^{2\mu}\,,\qquad
\phi_0'\sim B_\phi e^{-\Delta\mu}\,,\qquad \mu\to\infty\,.
\label{sm:g:bdy_falloff}
\end{equation}
The theorem includes the normalizable branch in standard quantization with $\Delta>1$, including the marginal case $\Delta=2$, and the branch at the Breitenlohner--Freedman bound $\Delta=1$ with vanishing logarithmic source.
The other end is fixed by reflection.
The source preserving solution of the master equation has
\begin{equation}
\zz=O(e^{-2\Delta\mu})\,,\qquad
\eta=\frac{\zz}{f\Phi}=O(e^{-2\mu})\,.
\label{sm:g:bdy_decaying}
\end{equation}
To identify the excluded mode, let $Z_s(v)$ denote the constant leading coefficient of the second radial solution for $\zz$.
For $\Delta\ne1$, the corresponding $\eta\propto Z_s e^{2(\Delta-1)\mu}$ changes the coefficient of the independent scalar falloff $e^{-(2-\Delta)\mu_\text{FG}}$ after transformation to a Fefferman--Graham coordinate $\mu_\text{FG}$.
For $\Delta=2$ this changes the marginal boundary coupling.
At $\Delta=1$, this mode instead has asymptotically constant $\eta$ and generates a scalar term proportional to $Z_s\mu_\text{FG} e^{-\mu_\text{FG}}$.
Thus fixing the scalar source, or fixing the logarithmic source to zero at the BF bound, selects $Z_s=0$ and~\eqref{sm:g:bdy_decaying}.
These statements concern the independent asymptotic scalar modes; general mixed scalar boundary conditions and backgrounds maintained by spatially varying relevant sources are outside the theorem considered here.

On the selected branch the first-order radial coordinate shift can be written asymptotically as
\begin{equation}
\delta\mu_\text{FG}(\mu,v)=\xi_\infty(v)
-\int_\mu^\infty\eta(\mu',v)\,d\mu'\,,
\label{sm:g:bdy_radial_shift}
\end{equation}
where $\xi_\infty$ implements the boundary conformal representative.
The integral converges by~\eqref{sm:g:bdy_decaying}.
It rescales the normalizable scalar coefficient without changing the fixed source.
The remaining boundary conditions concern physical geometry.
We hold fixed the conformal length of each boundary arc and require the variation $\delta\ln N=h_\tau$ to remain bounded toward its interface endpoints.
Allowed boundary coordinate transformations preserve those endpoints.
In the present coordinates, the condition on arc lengths is
\begin{equation}
\delta\ell_\text{arc}
=\int_{-\infty}^{\infty}dv\,\operatorname{sech}v\,
   (h_{v,\infty}-h_{\tau,\infty})
=\left[\operatorname{sech}v\,\partial_vE_\infty\right]_{-\infty}^{\infty}
=0\,.
\label{sm:g:bdy_arc_length}
\end{equation}
The final equality is a boundary condition, not an identity for an arbitrary perturbation.
It is insufficient on its own: an angular homogeneous mode can shift $h_{v,\infty}$ and $h_{\tau,\infty}$ equally while leaving this difference unchanged.
Boundedness of the physical lapse excludes that remaining mode.

With standard holographic renormalization, the on-shell boundary variation has the form
\begin{equation}
\delta I_\text{ren}\big|_{\partial M}
=\frac12\int\sqrt{g_{(0)}}\,T^{ij}\delta g_{(0)ij}
 +\int\sqrt{g_{(0)}}\,\langle\mathcal O\rangle\delta J\,.
\label{sm:renormalized_variation}
\end{equation}
At the BF bound, $J$ is the coefficient of the logarithmic scalar falloff.
The standard counterterm prescription with $J=0$~\cite{Skenderis:2002wp} therefore gives no additional scalar boundary variation when the replica index changes at fixed sources.
The cosmic brane relation in~\eqref{sm:g:replica} remains valid on this branch, even though the bare geometric mass need not equal the renormalized boundary energy.
In particular, regularity of the lapse at the tip gives
\begin{equation}
\frac{\ccap}{\ceff}
=-\partial_n\ln N_*(n)\big|_{n=1}
=-h_\tau(0,0)\,.
\label{sm:tip_capacity}
\end{equation}

\subsection{Boundary Einstein equation}
The scalar equation in the uniform scalar gauge gives
\begin{equation}
\chi'=\frac{f'}{f}\eta+\frac{\zz'}{f\Phi}\,,
\qquad \chi=2\psi+\Dt E\,.
\label{sm:g:bdy_sigma}
\end{equation}
Substituting this relation into the fourth Einstein combination of Section~\ref{sec:two} gives
\begin{equation}
(2-\Dt)\psi
=\zz+\frac{f'}{2f}\frac{\zz'}{\Phi}\,,
\qquad
(2-\Dt)\psi_\infty=\zz_\infty-j\,,
\qquad
j(v)\equiv-\lim_{\mu\to\infty}\frac{\zz'}{\Phi}\,.
\label{sm:g:bdy}
\end{equation}
The limits are finite on the branch \eqref{sm:g:bdy_decaying}, where $\zz_\infty=0$.
Introduce the even angular Green function
\begin{equation}
w(v)=1-\sinh|v|\,\operatorname{arccot}(\sinh|v|)\,,
\qquad
(2-\Dt)w=\pi\delta(v)\,.
\label{sm:g:bdy_angular_green}
\end{equation}
It decays at both interface endpoints.
The second even solution is
\begin{equation}
H_0(v)=1+\sinh v\,\arctan(\sinh v)\,,\qquad
(2-\Dt)H_0=0\,.
\label{sm:angular_homogeneous}
\end{equation}
It grows as $\pi|\sinh v|/2$.
The two physical boundary conditions just stated remove its coefficient in $\psi_\infty$.
Consequently
\begin{equation}
h_{\tau,\infty}(0)=\psi_\infty(0)
=\frac1\pi\int_{-\infty}^{\infty}dv\,
\cosh v\,w(v)\,[\zz_\infty(v)-j(v)]\,.
\label{sm:g:bdy_lapse}
\end{equation}

\subsection{Why the transport term appears}
The physical metric is reflection symmetric across $\mu=0$ away from the tip.
The auxiliary split into $\psi$ and $E$ is nonlocal in $v$, however, and its pieces have jumps along that entire slice.
Differentiating $(2-\Dt)\Psi=2\zz$ and using $\Psi'=-2f\Phi E'$ gives
\begin{equation}
(2-\Dt)E'(0^+,v)=\frac{\pi}{\sst}\delta(v)\,,
\qquad
(t_v\partial_v-1)E'(0^+,v)=0\quad(v\ne0)\,.
\label{sm:auxiliary_tip_conditions}
\end{equation}
The second condition is precisely $h_\tau'(0^+,v)=0$.
Reflection symmetry assigns equal and opposite $\mu$-directed fluxes to the two sides of the conical source and makes the physical lapse derivative vanish away from the tip; without it, the source fixes only the flux difference.
Since $(t_v\partial_v-1)w=(t_v\partial_v-1)H_0=-\operatorname{sech}^2v$, these conditions select
\begin{equation}
E'(0^\pm,v)=\pm\frac{w(v)-H_0(v)}{\sst}
=\mp\frac{\pi}{2\sst}\sinh|v|\,.
\label{sm:auxiliary_tip_growth}
\end{equation}
This growth is required in the auxiliary fields; it cancels in the physical perturbations $h_v$ and $h_\tau$.
Imposing decay separately on $E'$ or $\psi'$ would discard the response that fixes the capacity normalization.

To track that response invariantly, define the conformal length variation of a slice of constant scalar field,
\begin{equation}
\mathcal A(\mu)
=\int_{-\infty}^{\infty}dv\,\operatorname{sech}v\,[h_v-h_\tau]
=\left[\operatorname{sech}v\,\partial_vE\right]_{-\infty}^{\infty}\,.
\label{sm:slice_arc}
\end{equation}
The equation $(fE')'=-\eta$, together with the decay of $\eta$ at the endpoints, implies $(f\mathcal A')'=0$.
Equation~\eqref{sm:auxiliary_tip_growth} fixes $f\mathcal A'=-\pi\sst$, and the fixed boundary arc gives $\mathcal A(\infty)=0$.
Thus
\begin{equation}
\mathcal A(\mu)=\pi\sst\int_\mu^\infty\frac{d\mu'}{f(\mu')}\,.
\label{sm:slice_arc_solution}
\end{equation}
The conformal length of an interior slice is not fixed.
For pure AdS with a probe scalar, this reduces to $\pi(1-\tanh\mu)$, the first-order change in the exact geometry with a conical defect.

Boundedness of the physical $h_\tau=\psi+t_v\partial_vE$ fixes the coefficient of $H_0$ in $\psi(\mu,v)$ to be $-\mathcal A(\mu)/\pi$.
At the tip slice $f'(0)=0$, so~\eqref{sm:g:bdy} reduces to $(2-\Dt)\psi(0,v)=\zz(0,v)$.
Its value at $v=0$, including the homogeneous part, gives the direct response formula
\begin{equation}
\frac{\ccap}{\ceff}
=\sst B-\frac1\pi\int_{-\infty}^{\infty}dv\,
\cosh v\,w(v)\zz(0,v)\,.
\label{sm:direct_tip_response}
\end{equation}
The first term is fixed by the change in conformal length sourced by the tip.
The second is the scalar wall's relaxation, and its sign is fixed by the positive Green function of~\eqref{sm:master}.

\subsection{Cancellation of the boundary flux}
Reflection symmetry and the conical source in~\eqref{sm:master} imply
\begin{equation}
\frac{\zz'(0^+,v)}{\Phi(0)}=-\pi\sst\,\delta(v)\,.
\label{sm:g:bdy_tip_flux}
\end{equation}
Integrating the master equation over the radial half line and applying the Green function \eqref{sm:g:bdy_angular_green} gives
\begin{equation}
(2-\Dt)\int_0^\infty\!\eta(\mu,v)\,d\mu
=\pi\sst\,\delta(v)-j(v)\,,
\qquad
\int_0^\infty\!\eta(\mu,0)\,d\mu
=\sst-\frac1\pi\int_{-\infty}^{\infty}\!dv\,
\cosh v\,w(v)j(v)\,.
\label{sm:g:bdy_integrated_master}
\end{equation}
The axis equation \eqref{sm:g:Daxis} gives
\begin{equation}
h_{\tau,\infty}(0)-h_\tau(0,0)
=\int_0^\infty\!d\mu\,
\left[\frac{f'}{2f}+\int_\mu^\infty\frac{d\mu'}{f(\mu')}\right]
\eta(\mu,0)\,.
\label{sm:g:bdy_integrated_lapse}
\end{equation}
Define
\begin{equation}
\Xi_\Phi(\mu)=\int_\mu^\infty\Phi(\mu')\,d\mu'\,,
\qquad
\frac{f'}{2f}+\int_\mu^\infty\frac{d\mu'}{f(\mu')}
=1+\Xi_\Phi(\mu)\,.
\label{sm:g:bdy_Xi}
\end{equation}
The second identity follows because both sides have derivative $-\Phi$ and approach one at infinity.
Combining~\eqref{sm:g:bdy_lapse}, \eqref{sm:g:bdy_integrated_master}, and \eqref{sm:g:bdy_integrated_lapse} cancels the boundary flux $j$ and yields
\begin{equation}
\frac{\ccap}{\ceff}=-h_\tau(0,0)
=\sst+\int_0^\infty\!d\mu\,\Xi_\Phi(\mu)\eta(\mu,0)
-\frac1\pi\int_{-\infty}^{\infty}\!dv\,
\cosh v\,w(v)\zz_\infty(v)\,.
\label{sm:g:direct}
\end{equation}
The final term vanishes under the source condition \eqref{sm:g:bdy_decaying}.
Consequently
\begin{equation}
\frac{\ccap}{c}
=\sst^2+\sst\int_0^\infty\!d\mu\,
\Xi_\Phi(\mu)\eta(\mu,0)
=\sst^2\bigl[1+\pi U_\Phi(0,0)\bigr]\,,
\label{sm:g:SU}
\end{equation}
where $U_\Phi$ is the decaying solution of the master operator with source
$[\Xi_\Phi(|\mu|)/(f\Phi)]\delta(v)$.
The last equality follows by reciprocity with respect to the measure $\cosh v\,d\mu\,dv$ and the source normalization in~\eqref{sm:master}.
This derivation uses the physical lapse and boundary conditions directly, and therefore does not require identifying a bare geometric mass with the renormalized energy at slow scalar falloff.

\subsection{Partial wave representation}
Let
\begin{equation}
Y_p(v)={}_2F_1\!\left(\frac14+\frac{ip}{2},\frac14-\frac{ip}{2};
\frac12;-\sinh^2v\right)\,,
\qquad
\Dt Y_p=-\left(p^2+\frac14\right)Y_p\,,
\qquad k_p^2=p^2+\frac94\,.
\label{sm:g:bdy_partial_waves}
\end{equation}
The spectral density is
\begin{equation}
\rho(p)=\frac{p\sinh(\pi p)}{4\pi^3}
\left|\Gamma\!\left(\frac14+\frac{ip}{2}\right)\right|^4\,.
\label{sm:spectral_density}
\end{equation}
It is normalized so that
\begin{equation}
\delta(v)=\int_0^\infty dp\,\rho(p)Y_p(v)\,,\qquad
U_\Phi(0,0)=\int_0^\infty dp\,\rho(p)u_p(0)\,.
\label{sm:spectral_resolution}
\end{equation}
The radial function is
\begin{equation}
u_p(0)=\int_{-\infty}^{\infty}\!d\mu\,
G_p(0,\mu)\frac{\Xi_\Phi(|\mu|)}{f(\mu)\Phi(\mu)}\,,
\label{sm:g:bdy_radial_u}
\end{equation}
where the radial Green function is defined by
\begin{equation}
\mathcal L_pG_p(\mu,\nu)=\delta(\mu-\nu)\,,\qquad
\mathcal L_p=-\partial_\mu\Phi^{-1}\partial_\mu
 +\frac{k_p^2}{f\Phi}\,.
\label{sm:radial_green}
\end{equation}
It has the source preserving decay at both ends.
Its quadratic form is positive because $f>0$, $\Phi>0$, and $k_p^2>0$.
Applying its radial operator to $\Xi_\Phi(|\mu|)$ gives
\begin{equation}
\left[-\partial_\mu\Phi^{-1}\partial_\mu
      +\frac{k_p^2}{f\Phi}\right]\Xi_\Phi(|\mu|)
=2\delta(\mu)+\frac{k_p^2\Xi_\Phi(|\mu|)}{f\Phi}\,.
\label{sm:g:bdy_radial_identity}
\end{equation}
Both $G_p$ and $\Xi_\Phi$ have the decay required by fixed sources, so the Green boundary term vanishes and
\begin{equation}
u_p(0)=\frac{\Xi_\Phi(0)-2G_p(0,0)}{k_p^2}\,.
\label{sm:g:bdy_radial_resolvent}
\end{equation}
Finally,
\begin{equation}
\int_0^\infty\!dp\,\frac{\rho(p)}{k_p^2}
=\frac{w(0)}\pi=\frac1\pi\,,
\qquad
1+\Xi_\Phi(0)=\int_0^\infty\frac{d\mu}{f(\mu)}
=\frac{c}{\cLR}\,.
\label{sm:g:bdy_normalizations}
\end{equation}
Substitution into~\eqref{sm:g:SU} gives
\begin{equation}
\ccap=\frac{\ceff^2}{\cLR}
-\frac{2\pi\ceff^2}{c}
\int_0^\infty dp\,\frac{\rho(p)}{k_p^2}G_p(0,0)\,.
\label{sm:response}
\end{equation}
The same formula follows immediately from~\eqref{sm:direct_tip_response} and the partial wave expansion of $\zz$.
The flux calculation provides a second derivation in which all boundary terms remain explicit.
The identification of $\cLR$ uses the smooth wall transport formula of~\cite{Bachas:2020yxv,Bachas:2022etu}.

\section{Elementary estimates and thin branes}\label{sec:four}
With $g=G_p(\cdot,0)$, $G_p(0,0)^{-1}=\min_{g(0)=1}\int_{-\infty}^\infty(g'^2+k_p^2g^2/f)\,d\mu/\Phi$.
Positivity of $G_p$ gives the upper bound in~\eqref{sm:bound}.
Dropping the second term gives $G_p(0,0)\le\tfrac12\Xi_\Phi(0)$, that is $\ccap>\ceff^2/c$, which is weaker than \eqref{sm:bound}.
The inequality $g'^2+k^2g^2/f\ge k|(g^2)'|/\sqrt f$ gives $G_p(0,0)\le\max(\sqrt f\Phi)/(2k_p)$ and the estimate
\begin{equation}
\ccap\geq\frac{\ceff^2}{\cLR}-a_\text{sp}\,\frac{\ceff^2}{c}\max_\mu(\sqrt f\,\Phi)\,,
\qquad a_\text{sp}=\pi\int_0^\infty\frac{\rho(p)}{k_p^3}dp\leq\frac49\,.
\label{sm:g:throat}
\end{equation}
Numerically, $a_\text{sp}=\pi\int_0^\infty dp\,\rho(p)k_p^{-3}=0.40664$.
Conjugation by $\sqrt{\cosh v}$ maps $2-\Dt$ to $-\partial_v^2+\tfrac94+\tfrac14\text{sech}^2v\ge-\partial_v^2+\tfrac94$, and heat kernel comparison gives the analytic bound $a_\text{sp}\le\tfrac12\int_{-\infty}^\infty dk\,(k^2+\tfrac94)^{-3/2}=\tfrac49$.
With $z_\phi=1/\phi_0'$, $h=f'/(2f)$ and $\tilde g=g/\phi_0'$, the energy on the half line is $\int_0^\infty[\tilde g'^2+(z_\phi''/z_\phi+k_p^2/f)\,\tilde g^2]\,d\mu$.
If $z_\phi''/z_\phi-\tfrac12h'-\tfrac14h^2\ge0$, completing the square with $R_p=k_p/\sqrt f-\tfrac12h$, for which $R_p^2-R_p'=k_p^2/f+\tfrac12h'+\tfrac14h^2$, bounds it below by $R_p(0)\,\tilde g(0)^2$ and gives $G_p(0,0)\le\sqrt{f(0)}\,\Phi(0)/(2k_p)$, so that the maximum in~\eqref{sm:g:throat} may be replaced by the value at the brane.
The condition holds for Janus, where it reads $\frac{3f''}{4f}+\frac{3f'^2}{16f^2}\ge0$, and for the solvable wall of Section~\ref{sec:six}.

A useful sufficient condition for approaching the upper bound is $\Phi\to0$ uniformly in a neighborhood of the center whose width stays fixed, with $f$ uniformly bounded there.
In the dual principle of Section~\ref{sec:five} one can then choose a trial that decreases from one to zero inside this neighborhood with vanishing weighted energy.
This gives $G_p(0,0)\to0$ and $\ccap-\ceff^2/\cLR\to0$ under the corresponding bounded normalization.
Pointwise smallness of $\Phi(0)$ alone does not provide this conclusion.
A nontrivial solution with a smooth potential cannot have $\Phi=0$ on an open interval, by uniqueness of the scalar background equation.
Thin brane limits with a vacuum core provide examples approaching equality.
In that limit $U_\Phi=\Xi_\Phi(0)\,w(v)/\pi$ in the central region, with $w$ defined in Section~\ref{sec:three}.
The symmetric construction with two branes of~\cite{Baig:2022cnb} contains a central AdS region of radius $L_C=\sst$.
For equal wall tensions with $\int\Phi\,d\mu=8\pi G\sigma$ across each wall, the result is $\ccap/\ceff=(1+8\pi G\sigma)L_C$, in agreement with~(12) and~(13) of~\cite{Kusuki:2026arj} for equal outer radii and tensions.
For those formulas in general, with $x_i\equiv L_i/(1+8\pi GL_i\sigma_i)$, one finds $\ccap\cLR/\ceff^2=4x_Lx_R/(x_L+x_R)^2\le1$.

\section{Bounding the scalar relaxation}\label{sec:five}
The upper bound in~\eqref{sm:bound} follows directly from $G_p(0,0)>0$.
The lower bound requires an upper estimate on that subtraction.
We obtain it from a trial function whose angular energy can be calculated exactly.
The background Einstein equation supplies precisely the geometric weight against which that energy must be compared.

\subsection{Dual variational principle}
Fix $p$ and write $g(\mu)=G_p(\mu,0)$ on the radial half line.
An admissible trial $\alpha$ is locally absolutely continuous, has $\alpha(0)=1$, has finite weighted energy, and obeys $\alpha g\to0$ at infinity.
Integration by parts and Cauchy--Schwarz give
\begin{align}
g(0)&=-\int_0^\infty(\alpha g)'\,d\mu\,,
\nonumber\\
g(0)^2&\le
\left[\int_0^\infty\Phi
 \left(\alpha^2+\frac{f\alpha'^2}{k_p^2}\right)d\mu\right]
\left[\int_0^\infty\frac1\Phi
 \left(g'^2+\frac{k_p^2g^2}{f}\right)d\mu\right]\,.
\label{sm:dual_cauchy}
\end{align}
The second bracket is $g(0)/2$, by the Green equation and the reflection symmetry of its source jump.
Therefore
\begin{equation}
G_p(0,0)\le\frac12\int_0^\infty\Phi
\left(\alpha^2+\frac{f\alpha'^2}{k_p^2}\right)d\mu\,.
\label{sm:dual_inequality}
\end{equation}
Equality holds for
\begin{equation}
\alpha_p^\text{opt}(\mu)=-\frac2\Phi\partial_\mu G_p(\mu,0)\,.
\label{sm:optimal_trial}
\end{equation}
The optimal trial need not vanish at infinity; the condition on the boundary product and finite weighted energy are the relevant requirements.
Summing with the positive spectral weight yields
\begin{align}
\frac{\ccap}{c\sst^2}
 &=B-\min_\alpha\int_0^\infty d\mu\,\Phi(\mu)\cK_\alpha(\mu)\,,
\nonumber\\
\cK_\alpha
 &=\pi\int_0^\infty dp\,\frac{\rho(p)}{k_p^2}
 \left[\alpha_p^2+\frac{f\alpha_p'^2}{k_p^2}\right]\,.
\label{sm:g:Sdual}
\end{align}

\subsection{Available geometric weight}
Introduce two background quantities,
\begin{equation}
\mathcal C=A'^2+e^{-2A}\,,\qquad
T=\frac{A'}{\sqrt{\mathcal C}}\,.
\label{sm:geometric_weight}
\end{equation}
The Einstein equation implies $\mathcal C'=-2A'\Phi$.
Since $\sqrt{\mathcal C(0)}=1/\sst$ and $\sqrt{\mathcal C(\infty)}=1$, integration gives
\begin{equation}
\frac1{\sst}-1=\int_0^\infty T\Phi\,d\mu\,,\qquad
B-\frac1{\sst}=\int_0^\infty\Phi(1-T)\,d\mu\,.
\label{sm:available_weight}
\end{equation}
The second identity recovers the holographic transmission--entropy bound $\cLR\le\ceff$~\cite{Karch:2024udk}.
It remains valid when $f$ has additional extrema, because $|T|<1$ and hence $1-T>0$ everywhere.
To establish the capacity lower bound it is enough to construct an admissible trial with $\cK_\alpha\le1-T$ pointwise.

\subsection{An exactly integrable angular kernel}
For an even function $F(v)$, our spectral convention is
\begin{equation}
F(v)=\int_0^\infty dp\,\rho(p)F_pY_p(v)\,,\qquad
F_p=\int_{-\infty}^{\infty}dv\,\cosh v\,F(v)Y_p(v)\,.
\label{sm:parseval_convention}
\end{equation}
Parseval's identities then read
\begin{align}
\int_{-\infty}^{\infty}dv\,\cosh v\,F^2
 &=\int_0^\infty dp\,\rho(p)F_p^2\,,
\nonumber\\
\int_{-\infty}^{\infty}dv\,\cosh v\,F(2-\Dt)F
 &=\int_0^\infty dp\,\rho(p)k_p^2F_p^2\,.
\label{sm:parseval_energy}
\end{align}
Let $y=\sinh v$, so that $\cosh v\,dv=dy$ and $2-\Dt=2-\partial_y(1+y^2)\partial_y$, and let $z=\tanh m=\sin x$.
Define
\begin{equation}
w_z(v)=\int_z^1\frac{u^2\,du}{\sinh^2v+u^2}
=1-z-|y|\left[\arctan\frac1{|y|}-\arctan\frac z{|y|}\right]\,.
\label{sm:comparison_kernel}
\end{equation}
At $z=0$ it reduces to the decaying angular Green function $w(v)$ of Section~\ref{sec:three}.
At $z=1$ it vanishes.
Since $(2-\Dt)[u^2/(y^2+u^2)]=-\partial_u[2u^3(1-u^2)/(y^2+u^2)^2]$,
\begin{equation}
(2-\Dt)\,w_z=\frac{2z^3(1-z^2)}{(y^2+z^2)^2}\,,\qquad \partial_xw_z=-\frac{z^2\sqrt{1-z^2}}{y^2+z^2}\,.
\label{sm:g:wz}
\end{equation}
Define $\hat\alpha_p(m)\equiv(k_p^2/\pi)(w_z)_p=\pi^{-1}\!\int dy\,Y_p\,(2-\Dt)w_z$.
As $z\to0$ the first function in~\eqref{sm:g:wz} tends to $\pi\delta(y)$, so $\hat\alpha_p(0)=1$, and as $z\to1$ it vanishes, so $\hat\alpha_p(\infty)=0$.
On pure AdS $f=\cosh^2m$ and $\cosh m\,\partial_m=\partial_x$, hence
\begin{equation}
\cK_{\hat\alpha}=\pi\!\int_0^\infty\!dp\,\frac{\rho}{k_p^2}\Big[\hat\alpha_p^2+\frac{(\partial_x\hat\alpha_p)^2}{k_p^2}\Big]=\frac1\pi\int_{-\infty}^\infty\!dy\,\big[w_z(2-\Dt)w_z+(\partial_xw_z)^2\big]\equiv Z(z)+H(z)\,.
\label{sm:kernel_energy}
\end{equation}
With $\int dy\,(y^2+z^2)^{-2}=\frac{\pi}{2z^3}$ and $\int dy\,u^2(y^2+u^2)^{-1}(y^2+z^2)^{-2}=\frac{\pi u(u+2z)}{2z^3(u+z)^2}$,
\begin{equation}
H(z)=\tfrac12z(1-z^2)\,,\qquad Z(z)=(1-z^2)\int_z^1\!du\,\frac{u(u+2z)}{(u+z)^2}=1-\tfrac32z+\tfrac12z^3\,.
\label{sm:g:ZD}
\end{equation}
Thus $Z+H=1-z$, the cancellation needed for the geometric comparison.
In closed form, $z\int dy\,Y_p/(y^2+z^2)$ obeys Legendre's equation of degree $-\tfrac12+ip$ in $z$, is regular at $z=1$ and equals $\pi$ at $z=0$; therefore $\int dy\,Y_p/(y^2+z^2)=\pi\,\mathfrak P_p(z)/z$ with $\mathfrak P_p(z)=P_{-1/2+ip}(z)/P_{-1/2+ip}(0)$, and
\begin{equation}
\hat\alpha_p=(1-z^2)\,(\mathfrak P_p-z\,\mathfrak P_p'),\qquad\partial_x\hat\alpha_p=-k_p^2\,z\sqrt{1-z^2}\;\mathfrak P_p\,.
\label{sm:legendre_trial}
\end{equation}
The two halves of the identity are $\pi\int dp\,\rho\,(1-z^2)^2(\mathfrak P_p-z\mathfrak P_p')^2/k_p^2=Z$ and $\pi\int dp\,\rho\,\mathfrak P_p^2=\frac{1}{2z}$.
These identities in position space establish the normalization and energy analytically; a spectral truncation is not needed in the proof.

\subsection{Transfer to a general wall}
The natural comparison coordinate is $\widetilde\mu=\operatorname{artanh}T$.
The background equations give
\begin{equation}
f=\frac{\cosh^2\widetilde\mu}{\mathcal C}\,,\qquad
\widetilde\mu'=\frac{\mathcal C-\Phi}{\sqrt{\mathcal C}}\,.
\label{sm:comparison_coordinate}
\end{equation}
This coordinate can decrease when the warp factor has additional extrema.
We therefore use its running maximum,
\begin{equation}
m(\mu)=\max_{0\le\nu\le\mu}\widetilde\mu(\nu)\,,\qquad
U(\mu)=\tanh m(\mu)\ge T(\mu)\,.
\label{sm:running_maximum}
\end{equation}
It is nondecreasing and locally Lipschitz, with $m(0)=0$.
Take $\alpha_p(\mu)=\widehat\alpha_p(m(\mu))$.
Where the running maximum increases, $m=\widetilde\mu$ and the Einstein equation bounds its rate:
\begin{equation}
0\le m'=\frac{\mathcal C-\Phi}{\sqrt{\mathcal C}}
\le\sqrt{\mathcal C}\,,\qquad
\upsilon\equiv\frac{m'}{\sqrt{\mathcal C}}=1-\frac\Phi{\mathcal C}\in[0,1]\,.
\label{sm:comparison_rate}
\end{equation}
There the derivative contribution is $f\alpha_p'^2=\upsilon^2(\partial_x\widehat\alpha_p)^2$.
On every plateau, $m'=0$ and this contribution vanishes; set $\upsilon=0$ there.
No relation between $f$ and $\cosh^2m/\mathcal C$ is needed on a plateau.
Reflection symmetry also makes the two radial quadratic forms identical, giving the normalization on the half line used in~\eqref{sm:dual_cauchy}.
An asymmetric extension would require joining unequal left and right response functions at the brane and controlling their combined subtraction; this matching problem remains open, rather than showing that reflection symmetry is necessary for the inequality.
The exact kernel energy consequently gives, almost everywhere,
\begin{equation}
\cK_\alpha=Z(U)+\upsilon^2H(U)\,,\qquad
(1-T)-\cK_\alpha=(U-T)+(1-\upsilon^2)H(U)\ge0\,.
\label{sm:pointwise_bound}
\end{equation}
Substituting this trial into~\eqref{sm:g:Sdual} and using~\eqref{sm:available_weight} yields
\begin{equation}
\frac{\ccap}{c\sst^2}
\ge\frac1{\sst}
+\int_0^\infty d\mu\,\Phi\,[(U-T)+(1-\upsilon^2)H(U)]
\ge\frac1{\sst}\,.
\label{sm:lower_bound_proved}
\end{equation}
This proves $\ccap\ge\ceff$.
Only the background Einstein equation and $\Phi\ge0$ enter this comparison.
No additional sign or convexity condition on $V$ is needed within the stated class.
If $\Phi>0$ the integrand is positive wherever $0<U<1$, which holds beyond the first point where $A'>0$, so the inequality $\ccap>\ceff$ is strict.

\subsection{Remainder}
The dual principle is the minimum of a quadratic form.
Writing the trial function as $\alpha_p=\alpha_p^\text{opt}+\delta\alpha_p$, with $\delta\alpha_p(0)=0$, the cross term vanishes by the equation obeyed by $\alpha_p^\text{opt}$, and
\begin{align}
\frac{\ccap-\ceff}{c\sst^2}
={}&\int_0^\infty d\mu\,\Phi\,[(U-T)+(1-\upsilon^2)H(U)]
\nonumber\\
&+\pi\int_0^\infty dp\,\frac{\rho(p)}{k_p^2}
 \int_0^\infty d\mu\,\Phi
 \left[\delta\alpha_p^2+\frac{f\delta\alpha_p'^2}{k_p^2}\right]\,,
\label{sm:positive_remainder}
\end{align}
in which every term is nonnegative.

\subsection{Admissibility and the singular comparison weight}
Let $d\nu=\pi\rho(p)k_p^{-2}dp$.
The source trace follows in the strong energy norm:
\begin{equation}
\|\hat\alpha(z)-1\|_{L^2(d\nu)}^2=\frac{z+z^3}{2}\longrightarrow0\,.
\label{sm:trial_source_trace}
\end{equation}
The derivative energy is $H(z)=z(1-z^2)/2$, which tends to zero at both endpoints.
As $z\to1$, $w_z(v)\sim(1-z)\operatorname{sech}^2v$; the trial therefore has the required radial decay.
The running maximum is locally Lipschitz, so its first weak derivative introduces no delta function at a plateau boundary.
These facts justify the trial in the domain of the quadratic form.

The kernel also satisfies
\begin{equation}
-\partial_x(W_*\partial_xw_z)+W_*(2-\Delta_2)w_z=0\,,
\qquad W_*(x)=\frac{1}{\sin^2x\cos x}\,.
\label{sm:formal_weight}
\end{equation}
This is a formal Euler--Lagrange equation for the comparison profile $\Phi_*(m)=\coth^2m$.
Its energy diverges at the center and it is not an admissible scalar wall or a minimizer with finite energy.
The proof uses its regular extension kernel as a trial, not this singular profile as a physical background.

\section{An exact wall with a scalar potential}\label{sec:six}
For $0<b<1$ and $\alpha_b=\sqrt{1-b^2}/b$, consider the potential $V=-2-\alpha_b^2\cos^2(\phi/\alpha_b)$.
It admits the exact wall $f=b^2\cosh^2\mu$, $\phi_0=\alpha_b\arctan(\sinh\mu)$, with $\Phi=\alpha_b^2\operatorname{sech}^2\mu$.
The scalar saturates the Breitenlohner--Freedman bound at both ends ($m^2=-1$, $\Delta=1$), where it has no logarithmic source.
Here $\sst=b$, $c/\cLR=b^{-2}$, and $z_\phi''/z_\phi=1$ for $z_\phi=1/\phi_0'$, so with $\tilde g=G_p(\cdot,0)/\phi_0'$~\eqref{sm:radial_green} is $[-\partial_\mu^2+1+(k_p^2/b^2)\,\text{sech}^2\mu]\,\tilde g=\phi_0'(0)\,\delta(\mu)$.
Its decaying solution is an associated Legendre function of degree $-\tfrac12+i\nu_p$ and order minus one.
The diagonal Green function is
\begin{equation}
G_p(0,0)=\frac{\alpha_b^2}{4}\left|\frac{\Gamma(\tfrac34+i\nu_p/2)}{\Gamma(\tfrac54+i\nu_p/2)}\right|^2\,,\qquad \nu_p^2=\frac{k_p^2}{b^2}-\frac14\,.
\label{sm:solvable_green}
\end{equation}
It gives
\begin{equation}
\frac{\ccap}{c}=1-\frac{\pi(1-b^2)}{2}\int_0^\infty\!dp\,\frac{\rho(p)}{k_p^2}\,\Big|\frac{\Gamma(\tfrac34+\tfrac{i\nu_p}{2})}{\Gamma(\tfrac54+\tfrac{i\nu_p}{2})}\Big|^2,\qquad\nu_p^2=\frac{k_p^2}{b^2}-\frac14\,.
\label{sm:solvable_capacity}
\end{equation}
With $d_p=k_p/b-1$ and $\widehat R_p=1+d_p\,\text{sech}^2\mu$ one has $1+(k_p^2/b^2)\text{sech}^2\mu-(\widehat R_p^2-\widehat R_p')=\text{sech}^2\mu\,(d_p\tanh\mu-1)^2\ge0$, so the energy on the half line is at least $\widehat R_p(0)\,\tilde g(0)^2=(k_p/b)\,\tilde g(0)^2$ and $G_p(0,0)\le\alpha_b^2b/(2k_p)$.
With $\pi\int dp\,\rho k_p^{-3}\le\tfrac49$ this gives~\eqref{sm:solvable_bound}; the upper bound is $\ceff^2/\cLR=c$.
Explicitly,
\begin{equation}
\frac{\ceff}{c}=b\,,\qquad
\frac{\cLR}{c}=b^2\,,\qquad
1-\frac49b(1-b^2)<\frac{\ccap}{c}<1\,.
\label{sm:solvable_bound}
\end{equation}
Thus $\ccap/c\to1$ as $b\to0$, while the normalized entropy and transmission coefficients tend to zero.
The physical interpretation of the $b\to0$ endpoint is discussed in Section~\ref{sec:nine}.
Table~\ref{sm:tab:solvable} evaluates the exact integral and its analytic lower estimate.
Numerical minimization of the exact integral gives a minimum $\ccap/c$ of $0.8472$, at $b=0.567$, where $\ccap/\ceff=1.49$.

\begin{table}[h]
\begin{tabular}{cccc}
$b=\ceff/c$ & $\ccap/c$ (closed form) & $1-\tfrac49b(1-b^2)$ & $\ccap/\ceff$\\
\hline
0.9 & 0.934817639 & 0.924000 & 1.0387\\
0.7 & 0.860123048 & 0.841333 & 1.2287\\
0.5 & 0.850227191 & 0.833333 & 1.7005\\
0.3 & 0.889681048 & 0.878667 & 2.9656\\
0.1 & 0.959770524 & 0.956000 & 9.5977\\
\end{tabular}
\caption{\textsf{
Solvable wall: numerical quadrature of the exact spectral formula and the analytic lower bound.}}
\label{sm:tab:solvable}
\end{table}

\section{Janus and its long throat}\label{sec:seven}

The Janus background has
\begin{equation}
f(\mu)=\frac{1+\kappa\cosh2\mu}{2}\,,\qquad
\kappa=\sqrt{1-2\gamma^2}\,,\qquad
\phi_0'=\frac\gamma f\,,\qquad 0<\kappa<1\,.
\label{sm:janus_background}
\end{equation}
The scalar jump and the two background observables are
\begin{equation}
\Delta\phi=\sqrt2\operatorname{artanh}(\sqrt2\gamma)\,,\qquad
\frac{\ceff}{c}=\sqrt{\frac{1+\kappa}{2}}\,,\qquad
\frac{\cLR}{c}=\frac{\sqrt2\gamma}{\operatorname{artanh}(\sqrt2\gamma)}\,.
\label{sm:janus_observables}
\end{equation}
For $V=-2$, $\Phi=\gamma^2/f^2$, $\zz=\gamma^2\eta/f$ and~\eqref{sm:master} is $[\partial_\mu^2+f^{-1}\Dt-4]\eta=-(2\pi/\sst)\delta(\mu)\delta(v)$.
Equation~\eqref{sm:g:SU} can therefore be written as
\begin{align}
\frac{\ccap}{c}&=\sst^2+\pi\gamma^2U(0,0)\,,
\nonumber\\
\mathcal L U&\equiv[-\partial_\mu^2-f^{-1}\Delta_2+4]U
=\Xi(|\mu|)\delta(v)\,,\qquad
\Xi(\mu)=\int_\mu^\infty\frac{d\mu'}{f(\mu')^2}\,.
\label{sm:janus_auxiliary_response}
\end{align}
Here we used $\gamma^2=2\sst^2(1-\sst^2)$ and $\gamma^2\Xi(0)=\Delta\phi/(2\gamma)-1$.
The supersolution $U_+=f\,\Xi(|\mu|)\,w(v)/\pi$, which obeys $\cL U_+=\Xi\,\delta(v)+2w\,\delta(\mu)/(\pi f(0))$, gives the upper bound directly.

\subsection{Analytic lower bound}
In each partial wave the radial Green's function decreases when $f$ is replaced by its minimum, so
\begin{equation}
U(0,0)\ge U_\text{th}(\sst)=\int_0^\infty\!dp\,\frac{\rho(p)}{\kappa_p}\int_0^\infty\!d\mu\;\Xi(\mu)\,e^{-\kappa_p\mu}\,,\qquad \kappa_p^2=4+\frac{p^2+\tfrac14}{\sst^2}\,.
\label{sm:janus_comparison}
\end{equation}
Since $f\le\sst^2\cosh^2\!\mu$, $\Xi(\mu)\ge\sst^{-4}J(\mu)$ with $J=\int_\mu^\infty\text{sech}^4$.
Rescaling $\mu=\sst\nu$ and using that $J$ decreases and $\sst\kappa_p\le k\equiv\sqrt{p^2+17/4}$, every factor is at least its value at $\sst=1$, so $U(0,0)\ge\sst^{-2}U_\text{th}(1)$ with $U_\text{th}(1)=\int dp\,\rho(p)k^{-2}\int_0^\infty dt\,\text{sech}^4t\,(1-e^{-kt})$.
Because $k>2+\tfrac3{50}$, $e^{-x}\le1-x+\tfrac12x^2$, $\int_0^\infty\text{sech}^4t(1-e^{-2t})=\tfrac13$, $\int_0^\infty te^{-2t}\text{sech}^4t=\tfrac43\ln2-\tfrac56$ and $\int_0^\infty t^2e^{-2t}\text{sech}^4t<\tfrac14$, the inner integral exceeds $a_J\equiv\tfrac{17}{60}+\tfrac{2}{25}\ln2-\tfrac{9}{20000}$.
Conjugation by $\sqrt{\cosh v}$ maps $-\Dt$ to $H_v=-\partial_v^2+\tfrac14+\tfrac14\text{sech}^2v\le-\partial_v^2+\tfrac12$, so $\int dp\,\rho(p)/(p^2+\tfrac{17}{4})=(H_v+4)^{-1}(0,0)\ge\frac{1}{3\sqrt2}$.
Therefore $U_\text{th}(1)\ge \frac{a_J}{3\sqrt2}>\frac{1}{4\pi}$, since $\frac{4\pi a_J}{3\sqrt2}=1.0021$, and $\ccap/c>\sst^2+\frac{\gamma^2}{4\sst^2}=\tfrac12(1+\sst^2)$, which proves
\begin{equation}
\ccap>\frac{c^2+\ceff^2}{2c}>\ceff\,,
\label{sm:g:janusbound}
\end{equation}
for nontrivial Janus.

\subsection{Numerical evaluation}
For the numerical evaluation we diagonalize a finite difference representation of $-\partial_v^2+1/4+\operatorname{sech}^2v/4$ and solve the resulting radial tridiagonal systems.
The resulting values are listed below and plotted in the Letter.
The normalization in pure AdS is $U(0,0)=1/(3\pi)$.
The expansion for weak Janus coupling is
\begin{equation}
\frac{\ccap}{c}=1-\frac{\gamma^2}{6}+0.08025\,\gamma^4+O(\gamma^6)\,.
\label{sm:g:small}
\end{equation}
The quartic coefficient is numerical.
\subsection{Independent quartic check}
An independent perturbation in the scalar jump tests the response beyond its leading matter correction without using the uniform scalar gauge or the master equation.
Use polar coordinates about the cosmic brane, with $\theta\sim\theta+2\pi$, and expand in $a_\phi=\Delta\phi/2$:
\begin{align}
ds^2&=dr^2+e^{2P}d\theta^2+e^{2Q}du^2\,,
\nonumber\\
P&=\log(\sinh r/n)+a_\phi^2p_2+a_\phi^4p_4+\cdots\,,
\nonumber\\
Q&=\log(\cosh r/n)+a_\phi^2q_2+a_\phi^4q_4+\cdots\,,
\qquad \phi=a_\phi F_1+a_\phi^3F_3+\cdots\,.
\label{sm:polar:expansion}
\end{align}
The metric corrections and their first radial derivatives vanish at $r=0$, fixing the local brane length to $1/n$ and its opening angle to $2\pi/n$.
This local normalization does not fix the physical boundary scale of $u$.
The radial Einstein equations and scalar equation are
\begin{align}
P_{rr}+P_r^2&=1-\tfrac12(\phi_r^2+e^{-2P}\phi_\theta^2)\,,
\qquad
Q_{rr}+Q_r^2=1-\tfrac12(\phi_r^2-e^{-2P}\phi_\theta^2)\,,
\nonumber\\
0&=\partial_r(e^{P+Q}\phi_r)+\partial_\theta(e^{Q-P}\phi_\theta)\,.
\label{sm:polar:equations}
\end{align}
We impose regularity at the tip and $F_{1,\infty}=\operatorname{sgn}(\cos\theta)$, integrate the quadratic metric equations, and solve one linear equation for $F_3$.
Writing $\omega_n=n^2\operatorname{csch}^2r$ and $\mathcal D_n=\partial_r^2+2\coth(2r)\partial_r+\omega_n\partial_\theta^2$, this equation is
\begin{equation}
\mathcal D_nF_3
=2\omega_np_2F_{1,\theta\theta}
 -(p_{2,r}+q_{2,r})F_{1,r}
 -\omega_n(q_{2,\theta}-p_{2,\theta})F_{1,\theta}\,,
\qquad \mathcal D_nF_1=0\,.
\label{sm:polar:cubic}
\end{equation}
The two remaining Einstein constraints are evaluated independently on the resulting fields.

The physical boundary circumference before rescaling is $W=\int_0^{2\pi}d\theta\,\lim_{r\to\infty}e^{P-Q}$.
Consequently $u_\text{phys}=2\pi u/W$, and the refined entropy coefficient is
\begin{equation}
\frac{\widetilde c(n)}c=\frac{W}{2\pi n}
=\frac{1+a_\phi^2w_2(n)+a_\phi^4w_4(n)+\cdots}{n}\,.
\label{sm:polar:period}
\end{equation}
For a smooth regulator $F_{1,\infty}=f_M(\theta)$, holding the physical source fixed also requires the boundary reparametrization
\begin{equation}
\theta_\text{phys}=\theta+a_\phi^2\vartheta_2(\theta)+\cdots\,,
\qquad
\vartheta_2'=p_{2,\infty}-q_{2,\infty}-w_2\,,
\qquad F_{3,\infty}=\vartheta_2 f_M'\,.
\label{sm:polar:fixed_source}
\end{equation}
Reflection symmetry fixes $\vartheta_2=0$ at the interface rays.
The correction therefore vanishes distributionally for the exact step, but must be retained at finite smoothing.
With angular averaging denoted by brackets, the integrated radial equations give
\begin{align}
w_2&=\langle p_{2,\infty}-q_{2,\infty}\rangle\,,
\nonumber\\
w_4&=\left\langle\frac12(p_{2,\infty}-q_{2,\infty})^2
 +\int_0^\infty dr\,\left[
 e^{-r}(\cosh r\,q_{2,r}^2-\sinh r\,p_{2,r}^2)
\right.\right.
\nonumber\\
&\hspace{32mm}\left.\left.
 +e^{-2r}F_{1,r}F_{3,r}
 -\omega_nF_{1,\theta}F_{3,\theta}
 +\omega_np_2F_{1,\theta}^2\right]\right\rangle\,.
\label{sm:polar:quartic_functional}
\end{align}
The known entropy supplies $w_2(1)=-1/4$ and $w_4(1)=17/96$, while the quadratic replica result gives $w_2'(1)=-1/12$.
Using $a_\phi=\gamma+2\gamma^3/3+O(\gamma^5)$ and $\ccap=-\widetilde c'(1)$ yields
\begin{equation}
\frac{\ccap}{c}
=1-\frac{\gamma^2}{6}
 +\left[\frac{17}{96}-\frac29-w_4'(1)\right]\gamma^4
 +O(\gamma^6)\,.
\label{sm:polar:conversion}
\end{equation}
The polar calculation estimates the coefficient of $\gamma^4$ in the sharp Janus capacity by inserting the exact coefficients of the entropy and the quadratic response and differentiating the numerical quartic replica function:
\begin{equation}
c_4^\text{est}
=\frac{17}{96}-\frac29
 -\frac{w_4(1+h_n)-w_4(1-h_n)}{2h_n}\,,
\qquad h_n=0.00125\,.
\label{sm:polar:quartic_estimator}
\end{equation}
Table~\ref{sm:polar:convergence} gives this estimator for three source regulators at two resolutions and adds a third resolution for the hard Fourier regulator.
The hard regulator retains positive odd Fourier harmonics $m<M$ of $\operatorname{sgn}(\cos\theta)$, the Gaussian regulator multiplies the same Fourier coefficients by $e^{-(m/M)^2}$, and the third regulator is $\tanh(M\cos\theta)$.
All seven cases hold the physical regulated source fixed using~\eqref{sm:polar:fixed_source}, use radial extent $r_{\max}=10$, and evaluate both $n=1-h_n$ and $n=1+h_n$.
The fine calculation doubles the radial grid, angular grid, and source cutoff or steepness together, refining the mesh and making the source sharper at the same time.
The third hard Fourier calculation increases each of these fine-grid parameters by a factor of $3/2$.
\begin{table}[ht]
\centering
\begin{tabular}{lrrrrr}
\toprule
Regulator & $N_r$ & $N_\theta$ & $M$ & $c_4^\text{est}$ & $\varepsilon_4$\\
\midrule
Hard Fourier & 4800 & 2048 & 256 & 0.08037926 & $2.05\times10^{-6}$\\
Hard Fourier & 9600 & 4096 & 512 & 0.08031834 & $5.07\times10^{-7}$\\
Hard Fourier & 14400 & 6144 & 768 & 0.08029975 & $2.25\times10^{-7}$\\
Gaussian Fourier & 4800 & 2048 & 128 & 0.08047991 & $2.07\times10^{-6}$\\
Gaussian Fourier & 9600 & 4096 & 256 & 0.08038167 & $5.24\times10^{-7}$\\
Hyperbolic tangent & 4800 & 2048 & 128 & 0.08044000 & $2.09\times10^{-6}$\\
Hyperbolic tangent & 9600 & 4096 & 256 & 0.08035364 & $5.16\times10^{-7}$\\
\bottomrule
\end{tabular}
\caption{\textsf{
Independent polar estimates of the coefficient of $\gamma^4$ in $\ccap/c$.
Here $N_r$ is the number of radial intervals, $N_\theta$ the number of angular points, and $\varepsilon_4$ the larger root-mean-square residual of the two quartic Einstein constraints on $0.2<r<4$, maximized over the two replica indices.}}
\label{sm:polar:convergence}
\end{table}
The spread across regulators decreases from $1.01\times10^{-4}$ to $6.33\times10^{-5}$ under this refinement.
Each sequence decreases toward the value $0.08025$ obtained from the spectral response, with a remaining difference at finite resolution between $6.8\times10^{-5}$ and $1.3\times10^{-4}$ on the fine grid.
The third hard Fourier estimate differs from $0.08025$ by $4.97\times10^{-5}$.
Least-squares fits to the three hard Fourier estimates with $a+b/M$ and $a+b\log M/M$ give $a=0.0802592$ and $a=0.0802449$, respectively.
The corresponding maximum fit residuals are $7.92\times10^{-7}$ and $1.87\times10^{-6}$.
These extrapolations agree with the spectral value at the level of a few times $10^{-5}$ under the stated leading-error models.
The seven computations, fit analysis, and raw diagnostics are included in the reproduction archive.
The driver \texttt{run\_quartic\_convergence.py} runs the original six cases by default; \texttt{--resolution third --regulators hard} selects the additional case.
Section~\ref{sec:nine} specifies the interpretation and remaining errors of this estimator.

\subsection{The limit of a long throat}
The limit of a long throat can be taken directly in the radial Green function.
For every fixed $\mu$ as $\kappa\to0$,
\begin{equation}
f\to\frac12\,,\qquad \Phi\to2\,,\qquad
\mathcal L_p\to-\frac12\partial_\mu^2+k_p^2\,.
\label{sm:janus_throat_operator}
\end{equation}
The asymptotic AdS ends recede to infinite proper radial distance.
For Janus, conjugate the Green function according to
\begin{equation}
G_p(\mu,\nu)=\phi_0'(\mu)\phi_0'(\nu)\widetilde G_p(\mu,\nu)\,.
\label{sm:janus_green_conjugation}
\end{equation}
The radial problem then takes the Schr\"odinger form
\begin{equation}
\widetilde{\mathcal L}_p=-\partial_\mu^2+4+\frac{p^2+1/4}{f(\mu)}\,.
\label{sm:janus_throat_schrodinger}
\end{equation}
Its potential is bounded below by four uniformly in $\kappa$.
For each fixed $p$, this gap suppresses the influence of the distant ends, while the coefficients converge on every compact radial interval.
The diagonal Green function therefore approaches that of the operator with constant coefficients on the full line:
\begin{equation}
G_p^\text{th}(\mu,0)=\frac{e^{-\sqrt2k_p|\mu|}}{\sqrt2k_p}\,,\qquad
G_p(0,0)\longrightarrow\frac1{\sqrt2k_p}\,.
\label{sm:janus_throat_green}
\end{equation}
One may equivalently first fix a compact radial interval, use convergence of its uniformly positive quadratic form, and then send its boundaries to infinity.
The elementary estimate of Section~\ref{sec:four} gives a uniform spectral majorant:
$\max_\mu(\sqrt f\Phi)=\gamma^2/f(0)^{3/2}\le\sqrt2$ and hence $G_p(0,0)\le1/(\sqrt2k_p)$.
Since $\rho(p)/k_p^3$ is integrable, dominated convergence applies to the subtraction in~\eqref{sm:response}.
Writing
\begin{equation}
a_\text{sp}\equiv\pi\int_0^\infty dp\,\frac{\rho(p)}{k_p^3}
=0.40663741886693849\ldots\,,
\label{sm:spectral_constant}
\end{equation}
we obtain
\begin{align}
\frac{\ccap}{c}
 &=\frac12\log\frac2\kappa-\frac{a_\text{sp}}{\sqrt2}+o(1)
\nonumber\\
 &=\frac{\Delta\phi}{2\sqrt2}
   -0.28753607636500675\ldots+o(1)\,.
\label{sm:janus_asymptotic}
\end{align}
The divergence comes from $\sst^2B$, while the scalar subtraction has a finite throat limit.

Section~\ref{sec:nine} specifies the order of limits and the range addressed by this linear response calculation.

The figure uses $\mu_{\max}=v_{\max}=10$, mesh spacing $h=0.0125$, and 256 radial quadrature nodes.
Halving $h$ at four representative jumps changes the capacity by at most $2.72\times10^{-5}c$.

\begin{table}[h]
\begin{tabular}{ccccccc}
$\kappa$ & $\gamma^2$ & $\Delta\phi$ & $\ceff/c$ & $\cLR/c$ & $\ccap/c$ & $\ceff^2/(c\cLR)$\\
\hline
0.99 & 0.0100 & 0.2008 & 0.99750 & 0.99333 & 0.99835 & 1.00168\\
0.95 & 0.0488 & 0.4568 & 0.98742 & 0.96661 & 0.99209 & 1.00868\\
0.90 & 0.0950 & 0.6606 & 0.97468 & 0.93309 & 0.98510 & 1.01812\\
0.80 & 0.1800 & 0.9803 & 0.94868 & 0.86562 & 0.97435 & 1.03972\\
0.70 & 0.2550 & 1.2666 & 0.92195 & 0.79740 & 0.96897 & 1.06596\\
0.60 & 0.3200 & 1.5537 & 0.89443 & 0.72819 & 0.97083 & 1.09861\\
0.50 & 0.3750 & 1.8625 & 0.86603 & 0.65760 & 0.98287 & 1.14052\\
0.40 & 0.4200 & 2.2158 & 0.83666 & 0.58496 & 1.01020 & 1.19666\\
0.30 & 0.4550 & 2.6500 & 0.80623 & 0.50909 & 1.06271 & 1.27679\\
0.20 & 0.4800 & 3.2420 & 0.77460 & 0.42740 & 1.16351 & 1.40382\\
0.10 & 0.4950 & 4.2331 & 0.74162 & 0.33241 & 1.38966 & 1.65457\\
0.03 & 0.4995 & 5.9390 & 0.71764 & 0.23802 & 1.88273 & 2.16371\\
0.01 & 0.5000 & 7.4929 & 0.71063 & 0.18873 & 2.39040 & 2.67577\\
\end{tabular}
\caption{\textsf{
Janus interface.
All coefficients are normalized by the common central charge.}}
\label{sm:tab:data}
\end{table}

\section{Weak interfaces in conformal perturbation theory}\label{sec:eight}
\label{sm:cft}

\subsection{Interface family}
We consider a smooth family of conformal interfaces through the identity interface of a unitary two-dimensional CFT.
The first-order deformation is a jump $q=\Delta\lambda$ of a spinless exactly marginal bulk coupling, with
\begin{equation}
 \delta I=\frac q2\int d^2z\,\operatorname{sgn}(\operatorname{Re}z)\,\mathcal O(z)\,,
 \qquad
 \langle\mathcal O(z)\mathcal O(0)\rangle
 =\frac{G_{\mathcal O}}{|z|^4}\,,
 \qquad X=G_{\mathcal O}q^2\,.
 \label{sm:cft:source}
\end{equation}
The metric $G_{\mathcal O}$ is evaluated at the midpoint of the two bulk couplings.
We assume an analytic interface prescription that is covariant under exchange and has no additional defect degrees of freedom or first-order localized deformation.
We regulate the coupling step by an interpolation of physical width $a$ that is symmetric under exchange.
At fixed $a$, we first remove the cutoff on operator collisions after subtracting the local counterterms.
We then take the limit of a sharp interface $a/\epsilon_\text{UV}\to0$, with the same physical prescription on every replica, before extracting the coefficient as $L/\epsilon_\text{UV}\to\infty$.
Equivalently, the result is an infrared interface coefficient measured on scales large compared with $a$.

A rotation by $\pi$ changes $q$ to $-q$ and exchanges the two half lines.
Purity and covariance of the prescription under exchange therefore make the nonzero entanglement spectrum and energy reflection even in $q$.
This symmetry justifies an $O(q^4)$ remainder in the quadratic results below.
It also holds for endpoints chosen symmetrically along a geodesic of the conformal manifold, with contact terms defined in midpoint normal coordinates.
Although exactly marginal three-point coefficients vanish at separated points, their distributions can contain connection contact terms~\cite{Balthazar:2022hzb}.
A coordinate expansion at fixed endpoints can consequently contain cubic connection terms; the even expansion here follows from the centered exchange symmetry.

\subsection{Replica kernel and regulator}
For positive integer $n$, the map $w=u+i\theta=\log z$ gives a cylinder with $\theta\sim\theta+2\pi n$ and length $\ell=\log(L/\epsilon_\text{UV})$.
The marginal source is $q\operatorname{sgn}(\cos\theta)/2$ and the two-point function on the transparent cylinder is
\begin{equation}
 \langle\mathcal O(w)\mathcal O(0)\rangle_n
 =\frac{G_{\mathcal O}}{|2n\sinh(w/2n)|^4}\,.
 \label{sm:cft:correlator}
\end{equation}
Expanding the square wave in positive odd harmonics gives $b_m^2=16/(\pi^2m^2)$, where all sums over $m$ below mean $m=1,3,5,\ldots$.
With a strip cutoff $|u|>\delta$, the quadratic contribution before the local rotational correction is
\begin{equation}
 \frac{\delta\log Z_n}{\ell}
 =\frac{2X}{n}\sum_{m\ \mathrm{odd}}\frac{I(nm;\delta/n)}{m^2}\,,
 \qquad
 I(p;\varepsilon)=\int_\varepsilon^\infty
 \frac{e^{-pu}(p\sinh u+\cosh u)}{\sinh^3u}\,du\,.
 \label{sm:cft:fourier}
\end{equation}
At fixed $p$ its expansion for a small cutoff is
\begin{align}
 I(p;\varepsilon)
 ={}&\frac{1}{2\varepsilon^2}-\frac16-\frac p2
 +\frac{p^2}{2}\left[
 \log(2\varepsilon)+\gamma_E-\frac12
 +\psi\left(1+\frac p2\right)\right]
 +O(\varepsilon)\,.
 \label{sm:cft:fourier-expansion}
\end{align}
The local terms polynomial in $p^2$ that depend on the cutoff are contact terms.
In particular, the strip constant $-1/6$ is absent with a rotational cutoff, for which
\begin{equation}
 \int_{\substack{u\in\mathbb R,\,-\pi<\theta\le\pi\\|w|>\varepsilon}}d^2w\,
 |2\sinh(w/2)|^{-4}
 =\frac{\pi}{\varepsilon^2}+O(\varepsilon^2)\,.
 \label{sm:cft:rotational-cutoff}
\end{equation}
The integral is over one fundamental cylinder with a disk removed about its coincident point.
This prescription preserves the central charge under a uniform exactly marginal deformation.
Keeping the same physical cutoff $\delta$ on every replica sets $\varepsilon=\delta/n$ in~\eqref{sm:cft:fourier} and produces the term $-\log n$ in the finite digamma difference.
Interface tension and other local source terms cancel between $\log Z_n$ and $n\log Z_1$; endpoint terms do not affect the coefficient of $\ell$.

The cutoff limit can also be taken after summing the harmonics, avoiding a termwise ultraviolet expansion.
Define $A(v)=\operatorname{arctanh}(e^{-v})$ and $B(v)=\operatorname{Li}_2(e^{-v})-\tfrac14\operatorname{Li}_2(e^{-2v})$.
The normalized strip result is the convergent integral
\begin{align}
 \frac{\delta F_\text{strip}(n)}{X\ell}
 =2\int_0^\infty dv\,\bigg\{
 &A(v)\left[\frac{1}{n\sinh^2(v/n)}
             -\frac{n}{\sinh^2v}\right]\nonumber\\
 {}+{}&B(v)\left[\frac{\cosh(v/n)}{n^2\sinh^3(v/n)}
             -\frac{n\cosh v}{\sinh^3v}\right]\bigg\}
 \nonumber\\
 ={}&-\mathcal J_\text{CFT}(n)-\frac{\pi^2}{24}(n^{-1}-n)\,.
 \label{sm:cft:summed-strip}
\end{align}
The rotational contact correction removes the last term.
The universal logarithmic coefficient is therefore
\begin{align}
 \frac{F(n)}{\ell}
 ={}&\frac c{12}(n^{-1}-n)-X\mathcal J_\text{CFT}(n)+O(q^4)+o(1)\,,
 \label{sm:cft:renyi}\\
 \mathcal J_\text{CFT}(n)
 ={}&\sum_{m\ \mathrm{odd}}\left\{
 \frac{1-n}{m}
 -n\left[\psi\left(1+\frac{nm}{2}\right)
         -\psi\left(1+\frac m2\right)-\log n\right]\right\}
 \nonumber\\
 ={}&\int_0^\infty
 \frac{\tanh(u/n)-n\tanh u}{e^{2u}-1}\,du\,.
 \label{sm:cft:kernel}
\end{align}
Here $F(n)=\log\operatorname{Tr}\rho_A^n$ and $o(1)$ denotes terms vanishing after division by $\ell$ at fixed coupling.
Related integrals at integer replica index occur in~\cite{Mollabashi:2021xsd}.

\subsection{Continuation and coefficients}
The geometric construction using a square wave above initially applies at integer $n$.
A useful representation of its continuation is
\begin{align}
 \mathfrak g(z)
 &\equiv\log z-\psi(1+z)+\frac{1}{2z}
 =\int_0^\infty h(t)e^{-zt}\,dt\,,
 \qquad h(t)=\frac12\coth\frac t2-\frac1t\,,
 \label{sm:cft:binet}\\
 \Sigma(n)
 &\equiv\sum_{m\ \mathrm{odd}}\mathfrak g(nm/2)
 =\int_0^\infty\frac{h(t)}{2\sinh(nt/2)}\,dt\,,
 \qquad
 \mathcal J_\text{CFT}(n)=n[\Sigma(n)-\Sigma(1)]\,.
 \label{sm:cft:sigma}
\end{align}
These expressions are holomorphic for $\operatorname{Re}n>0$.
For $\operatorname{Re}n\ge1$, they obey $|\Sigma(n)|\le\Sigma(1)=\log2/2$ and hence $|\mathcal J_\text{CFT}(n)|\le|n|\log2$.
Carlson's theorem gives uniqueness among holomorphic continuations in this growth class~\cite{Witten:2018xfj}.
The identification with the physical replica continuation uses the assumption stated in Section~\ref{sec:nine}.
The exact determinant for the compact boson independently gives the same continuation~\cite{Sakai:2008tt,Eisler:2012xry}.

Differentiating the convergent kernel yields
\begin{equation}
 \mathcal J_\text{CFT}'(1)=-\frac{\pi^2}{16}\,,\qquad
 \mathcal J_\text{CFT}''(1)=\frac{\pi^2}{24}\,,\qquad
 c_\text{eff}=c-\frac{3\pi^2}{8}X+O(q^4)\,,\qquad
 c_\text{cap}=c-\frac{\pi^2}{4}X+O(q^4)\,.
 \label{sm:cft:coefficients}
\end{equation}
For the physical displacement operator, the Ward identity for the stress tensor gives $C_D=(c-c_{LR})/\pi^2$~\cite{Meineri:2019ycm}.
Its leading term is $D=q\mathcal O$, so the centered family has $\mathcal R\equiv1-c_{LR}/c=\pi^2X/c+O(q^4)$~\cite{Komatsu:2025cai}.
For a nonzero tangent this gives
\begin{equation}
 \frac{c_\text{eff}}c=1-\frac38\mathcal R+O(\mathcal R^2)\,,\qquad
 \frac{c_\text{cap}}c=1-\frac14\mathcal R+O(\mathcal R^2)\,.
 \label{sm:cft:reflection}
\end{equation}

For weak Janus,~\eqref{sm:janus_observables} gives
\begin{equation}
\mathcal R=\frac23\gamma^2+O(\gamma^4)\,,\qquad
\frac{\ceff}{c}=1-\frac14\gamma^2+O(\gamma^4)\,,\qquad
\frac{\ccap}{c}=1-\frac16\gamma^2+O(\gamma^4)\,.
\label{sm:cft:janus_match}
\end{equation}
The final coefficient agrees with the independent holographic response in~\eqref{sm:g:small}.
This comparison tests the scalar subtraction at its first nonzero order, including the normalization of the physical scalar jump.

\subsection{Escort fluctuations at fixed replica index}
For $\rho_A(n)=\rho_A^n/\operatorname{Tr}\rho_A^n$, let $\widetilde S(n)$ be its entropy and $C(n)$ the variance of its own modular Hamiltonian.
Then
\begin{equation}
 C(n)-\widetilde S(n)=n^2F''(n)-F(n)+nF'(n)\,,\qquad
 \delta[C(n)-\widetilde S(n)]
 =-X\ell n^2[3\Sigma'(n)+n\Sigma''(n)]+O(q^4\ell)\,.
 \label{sm:cft:escort}
\end{equation}
The expansion in partial fractions
\begin{equation}
 \frac{h(t)}t=\frac12\sum_{k=1}^\infty
 \frac{1}{(t/2)^2+\pi^2k^2}\,,
 \label{sm:cft:partial-fractions}
\end{equation}
is decreasing in $t$.
Integration by parts therefore gives
\begin{equation}
 3\mathfrak g'(x)+x\mathfrak g''(x)
 =\int_0^\infty t[th'(t)-h(t)]e^{-xt}\,dt<0\,.
 \label{sm:cft:escort-positive}
\end{equation}
Every term contributing to $3\Sigma'+n\Sigma''$ has this sign, proving positivity of the quadratic correction at each fixed $n>0$.

\subsection{Localized deformations and mixed tangents}
Now allow a first-order localized deformation with couplings $g^a$ and two-point metric $H_{ab}$.
At the identity interface its operators $\widehat{\mathcal O}_a$ are ambient spinless primaries of weights $(1/2,1/2)$.
Set $Y=H_{ab}g^ag^b$ and, for several bulk marginal operators, $X=G_{ij}\Delta\lambda^i\Delta\lambda^j$.
We assume an analytic family of conformal fixed points with these local tangents, excluding rotations generated by conserved currents and redundant directions.
The bulk--line two-point function vanishes at the identity because the ambient primary dimensions differ.
Combining the bulk result with the line calculation of~\cite{Kusuki:2026arj} gives, for an overall small parameter $q$,
\begin{align}
 c_\text{eff}&=c-\frac{3\pi^2}{8}X-\frac{3\pi^2}{2}Y+O(q^3)\,,\nonumber\\
 c_\text{cap}&=c-\frac{\pi^2}{4}X+\pi^2Y+O(q^3)\,,\nonumber\\
 c_{LR}&=c-\pi^2X-2\pi^2Y+O(q^3)\,,\nonumber\\
 2c_\text{eff}+c_\text{cap}-c_{LR}&=2c+O(q^3)\,.
 \label{sm:cft:mixed}
\end{align}
The same coefficients imply
\begin{equation}
\ccap-\frac{\ceff^2}{\cLR}
=2\pi^2Y-\frac{\pi^2}{2}X+O(q^3)\,.
\label{sm:cft:mixed_upper}
\end{equation}
A pure bulk jump lowers the capacity initially, whereas a pure localized tangent raises it.
At this order the upper inequality for walls supported by a scalar is consistent with a mixed tangent only when $Y\le X/4$.
For permitted mixed tangents, the leading slope $dc_\text{cap}/d\mathcal R$ lies between $-c/4$ and $c/2$.
The limiting displacement decomposition of~\cite{Komatsu:2025cai} supports the two possible local sectors, while their quadratic additivity here follows from two-point orthogonality.

A marginal line coupling can scale as $g=O(q)$ on a branch of conformal fixed points, as illustrated by a beta function of the form $\beta_g=Aq^2+Bqg+Cg^2+\cdots$~\cite{Karch:2018uft}.
Such a branch is described by~\eqref{sm:cft:mixed}, not by the relation for a pure jump~\eqref{sm:cft:reflection}.

\subsection{Checks at finite jump and in free fields}
For the compact boson, write the transmission amplitude as $s$, so $\mathcal R=1-s^2$, and let $\omega(x)=\operatorname{arsinh}(\sinh x/s)$.
Its exact capacity is
\begin{align}
 \frac{c_\text{cap}}c
 &=\frac6{\pi^2}\int_0^\infty\frac{\omega(x)^2}{\sinh^2\omega(x)}\,dx
 \nonumber\\
 &=1-\frac14\mathcal R+
 \left(\frac3{4\pi^2}-\frac3{16}\right)\mathcal R^2
 +O(\mathcal R^3)\,.
 \label{sm:cft:boson-finite}
\end{align}
The nonzero $\mathcal R^2$ coefficient gives the leading nonlinear correction to~\eqref{sm:cft:reflection}.
The exact determinants also give
\begin{equation}
 \lim_{\mathcal R\to0}
 \frac{2c_\text{eff}+c_\text{cap}-c_{LR}-2c}{c\mathcal R^2}
 =-\frac38\quad\text{(boson)}\,,\qquad
 =-\frac34\quad\text{(fermion)}\,,
 \label{sm:cft:sum-rule-corrections}
\end{equation}
where each theory is normalized by its own central charge.

To test the upper bound for walls supported by a scalar against the free theories, set $E=c_\text{eff}/c$ and $Q=c_\text{cap}/c$.
For the boson, changing variables to $\omega$ gives the Jacobian
\begin{equation}
 j_s(\omega)=\frac{s\cosh\omega}{\sqrt{1+s^2\sinh^2\omega}}\,,
 \qquad s\le j_s(\omega)\le1\,.
 \label{sm:cft:boson-jacobian}
\end{equation}
The entropy and capacity are positive integrals against this Jacobian, with transparent values one, so $E_B\ge s$ and $Q_B\le1$.
Consequently $Q_Bs^2<E_B^2$ for $0<s<1$.

For a Majorana or Dirac fermion, the modular dispersion and its inverse are
\begin{equation}
 \omega(x)=\operatorname{arcosh}(\cosh x/s)\,,\qquad
 x(\omega)=\operatorname{arcosh}(s\cosh\omega)\,.
 \label{sm:cft:fermion-spectrum}
\end{equation}
With $a_F=12/\pi^2$ and $\omega_0=\operatorname{arcosh}(1/s)$, integration by parts gives
\begin{equation}
 E_F=a_F\int_{\omega_0}^\infty x\omega\operatorname{sech}^2\omega\,d\omega\,,
 \qquad
 Q_F=a_F\int_{\omega_0}^\infty
 \omega^2\operatorname{sech}^2\omega\,x'(\omega)\,d\omega\,.
 \label{sm:cft:fermion-integrals}
\end{equation}
Cauchy--Schwarz then implies
\begin{align}
 E_F^2
 &\le Q_Fa_F\int_{\omega_0}^\infty
 \frac{x^2\operatorname{sech}^2\omega}{x'}\,d\omega
 \nonumber\\
 &=Q_Fa_Fs^2\int_0^\infty
 \frac{x^2\operatorname{sech}^2x\sinh^2x}{\sinh^2x+1-s^2}\,dx
 <Q_Fs^2\,,
 \label{sm:cft:fermion-upper-violation}
\end{align}
where $a_F\int_0^\infty x^2\operatorname{sech}^2x\,dx=1$ was used.
Thus $c_\text{cap}c_{LR}>c_\text{eff}^2$ for the fermion at every $0<s<1$, while the boson obeys the opposite strict inequality.
Both transparent limits saturate it.

\section{Scope and limitations}\label{sec:nine}

\subsection{Classical theorem and replica branch}
The proof applies to the Einstein--scalar walls specified in Sections~\ref{sec:one} and~\ref{sec:three}, with equal radii, reflection symmetry, a central global minimum, and the stated scalar sources and endpoint conditions.
The requirement of a central minimum is an independent geometric assumption.
We assume that the relevant saddle with replica and reflection symmetry exists, dominates, and is differentiable at $n=1$.
If several extremal lines are degenerate, the central branch must remain the relevant differentiable saddle.
The positive operator in Section~\ref{sec:two} controls the coupled response used in the proof; it does not establish stability against all other fluctuations or determine saddle dominance.
Potential independence means that no convexity or sign condition on derivatives of $V$ enters within this class.
The argument uses the full Einstein--scalar constraints and the monotonic scalar gauge, and is not a theorem for arbitrary matter obeying a null energy condition, multiple scalars, or asymmetric interfaces.

\subsection{Logarithmic coefficients and the Janus limit}
The limiting results concern the leading classical coefficients of $\ell$ in the entropy and capacity.
Terms of order $\ell^0$ can depend on the interface parameter, so a bounded entropy coefficient alone does not establish bounded full regulated entropy at large jump.
For Janus, the classical large-$c$ limit is taken first.
At each fixed $\kappa>0$, we extract the logarithmic coefficient at fixed physical sources and differentiate at $n=1$ before taking $\kappa\to0$.
Small first-order variation of the brane length necessarily requires
\begin{equation}
|n-1|\,\frac{\ccap}{\ceff}\ll1\,,
\label{sm:janus_linear_window}
\end{equation}
which becomes $|n-1|\Delta\phi\ll1$ parametrically in the throat limit.
This is a necessary condition for a small response, rather than a sufficient uniform validity estimate.
Establishing a uniform neighborhood in $n$ would require bounds on higher replica derivatives.
The calculation does not determine the growth of quantum corrections as $\kappa\to0$ or compare the chosen branch with competing saddles.

\subsection{The singular endpoint of the family with a cosine potential}
Changing $b$ in Section~\ref{sec:six} changes the potential, while each replica derivative holds that potential fixed.
The solutions form a bottom-up classical family, with no top-down embedding assumed.
Their curvature is
\begin{equation}
R(\mu)=-6-2\alpha_b^2\operatorname{sech}^2\mu\,,\qquad
R(0)=-4-\frac2{b^2}\,.
\label{sm:solvable_curvature}
\end{equation}
The endpoint $b=0$ is singular.
Semiclassical control requires the central curvature radius to exceed the Planck length, giving $b/G\gg1$, equivalently $bc\gg1$; a gravitational completion can introduce further ultraviolet length scales.
Thus the endpoint formulas describe normalized classical coefficients with large $c$ taken first.
They do not describe an arbitrarily small $b$ prediction at fixed gravitational cutoff or a regular factorizing interface in one fixed theory.
For a factorizing interface both the entropy and capacity coefficients vanish.

\subsection{Numerical evidence}
The bounds follow from the analytic proof in Section~\ref{sec:five}.
The comparisons across meshes and source regulators in Section~\ref{sec:seven} provide convergence and consistency checks, rather than certified continuum error bounds.
The estimator $c_4^\text{est}$ inserts exact entropy and quadratic response data for a sharp interface while retaining a regulated numerical replica derivative; at finite regulator it is not the direct quartic capacity coefficient of a separate exact Janus geometry.
Table~\ref{sm:polar:convergence} establishes a consistent convergence trend toward the spectral value $0.08025$.
The three-point extrapolations in Section~\ref{sec:seven} depend on the assumed leading-error model.
The displayed digits record computed outputs rather than certified significant digits.
The comparison between resolutions changes the radial grid, angular grid, and source regulator jointly, so it does not isolate any one error source.
Remaining errors include the finite outer radius, finite difference step in replica index, angular truncation, quadrature near the tip, and extrapolation to a sharp source.
The limit of a sharp source can contain powers of $\log M$ divided by $M$.
Fits with three parameters interpolate the three hard Fourier observations exactly and provide no residual degrees of freedom for testing their form.
The regulator spread and the small interior constraint residuals diagnose convergence but do not bound the error in the final observable.

\subsection{CFT branch, continuation, and finite jumps}
The calculation for a pure jump assumes the analytic conformal branch defined in Section~\ref{sec:eight}, which is covariant under exchange, with no first-order localized deformation.
Exact marginality of the ambient operator alone does not establish the existence of this branch.
The formulas for mixed tangents describe permitted local deformations; they do not establish the existence of every mixing ratio in a given CFT or cover arbitrary singular limits of thin branes.
To identify the analytic kernel with the physical replica coefficient, we assume that the regulated perturbative expansion and the logarithmic scaling limit preserve the holomorphy and growth condition stated around~\eqref{sm:cft:sigma}.
Carlson uniqueness fixes the continuation within that class, while membership of the physical perturbative coefficient in the class remains an additional assumption.
Equation~\eqref{sm:cft:reflection} holds for a small jump, with explicit nonlinear corrections in~\eqref{sm:cft:boson-finite}; the escort result likewise holds perturbatively at each fixed $n>0$, without claiming uniformity at finite jump as $n$ approaches zero or infinity.
The upper bound for walls supported by a scalar is not a general CFT inequality, as the exact fermion result in~\eqref{sm:cft:fermion-upper-violation} demonstrates.

\makeatletter

\immediate\write\@auxout{\string\citation{supplementStyle}}
\makeatother

\bibliographystyle{apsrev4-2}
\bibliography{refs}